\documentclass[times,twocolumn,twocolappendix,usenames,dvipsnames]{aastex631}

\usepackage{newtxtext,newtxmath}
\usepackage{amsmath}
\usepackage{comment}

\definecolor{dullred}{rgb}{0.7, 0.2, 0.2}
\definecolor{navyblue}{RGB}{0, 0, 128}
\usepackage{ulem}
\hypersetup{
    colorlinks=true,
    linkcolor=blue,
    filecolor=blue,      
    urlcolor=blue,
    citecolor=blue
}

\begin{document}

\title{A Non-parametric Method for the Inference of Halo Occupation Distributions}

\correspondingauthor{Jacob Kennedy}
\email{jnk97@physics.rutgers.edu}

\author[0009-0001-4745-3555]{Jacob Kennedy}
\affiliation{Department of Physics and Astronomy, Rutgers, The State University of New Jersey, Piscataway, NJ 08854, USA}

\author[0000-0003-1530-8713]{Eric Gawiser}
\affiliation{Department of Physics and Astronomy, Rutgers, The State University of New Jersey, Piscataway, NJ 08854, USA}
\affiliation{School of Natural Sciences, Institute for Advanced Study, Princeton, NJ 08540, USA}

\author[0000-0001-9298-3523]{Kartheik G. Iyer}
\affiliation{Columbia Astrophysics Laboratory, Columbia University, 550 West 120th Street, New York, NY 10027, USA}
\affiliation{Center for Computational Astrophysics, Flatiron Institute, 162 5th Ave, New York, NY 10010, USA}

\author[0000-0003-3466-035X]{L. Y. Aaron Yung}
\affiliation{Space Telescope Science Institute, 3700 San Martin Drive, Baltimore, MD 21218, USA}

\begin{abstract}
The galaxy-halo connection traces processes by which galaxies form and evolve. The halo occupation distribution (HOD) describes the relationship between galaxies and their host dark matter haloes. Measurements of the galaxy two-point correlation function (2PCF) allow us to extract information about the HODs of observed galaxy samples. Several parametric HOD models have been proposed in the literature, but the choice of parameterization restricts the space of possible HODs. To resolve this issue, we introduce a non-parametric HOD fitting method in which we train an emulator to learn the mappings among the galaxy 2PCF, physical properties used to select galaxy samples, and the HOD, all obtained from simulated past lightcones constructed with the Santa Cruz semi-analytic models. Implementing this emulator within a likelihood analysis framework, we derive constraints on the HOD of a galaxy sample when provided with a measurement of its 2PCF. Using the emulator to accelerate likelihood evaluations, we test the non-parametric HOD approach on a set of 2PCFs for mock galaxy samples drawn from the TNG100-1 simulation and selected above threshold values of stellar mass and star formation rate. Our framework is able to recover TNG100-1 HODs within 0.2 dex. We use the TNG100-1 mocks to tune the reported uncertainties to estimate those expected in the analysis of observations. Comparing to parametric HOD modelling routines applied to the same mock galaxy samples, our approach consistently infers the HOD with comparable or greater precision and accuracy.
\end{abstract}


\section{Introduction} \label{sec:intro}

The large-scale structure of the Universe describes how galaxies are distributed on distance scales greater than a single galaxy. While the distribution of galaxies appears homogeneous on very large distance scales (e.g.\ \citealt{Tegmark_2004, Hogg_2005, 10.1111/j.1365-2966.2005.09578.x}), deviations from this homogeneity on smaller scales encode a richness of information regarding the cosmological parameters of the Universe and the physics of galaxy formation and evolution. Measuring the statistical properties of the spatial distribution of galaxies has been a key focus of astronomical observations for the past several decades. Large scale galaxy surveys (e.g.\ Sloan Digital Sky Survey (SDSS); \citealt{2000AJ....120.1579Y}, Two Degree Field Galaxy Redshift Survey (2dFGRS); \citealt{2001MNRAS.328.1039C}, , Kilo-Degree Survey (KiDS); \citealt{KiDS2012}, DEEP2 Galaxy Redshift Survey (DEEP2); \citealt{Newman_2013}, Dark Energy Survey (DES); \citealt{2016MNRAS.460.1270D}) have provided a highly informative test-bed for candidate cosmological and galaxy formation theories. 

The statistic of merit in these analyses is often the galaxy two-point correlation function (2PCF; \citealt{1980lssu.book.....P}), which measures the degree of galaxy clustering as a function of scale. Cosmological inference from galaxy clustering is based on the assumption that galaxies trace the underlying dark matter distribution, which can be specified from initial conditions and simulated reliably and efficiently over large volumes for a given cosmological model \citep{2009ApJS..182..543A}. The prevailing cosmological paradigm predicts that all galaxies form and evolve within dark matter haloes. In this picture, haloes are defined as roughly spherical, virialized regions with a density $\approx 200$ times the average background density of the Universe (e.g.\ \citealt{1996MNRAS.281..703E, Cole_1996, 1996ApJ...462..563N}). 

One approach to modelling the galaxy-halo connection that has gained considerable traction is the halo occupation distribution (HOD; e.g.\ \citealt{2000ApJ...543..503M, 2000MNRAS.318.1144P, 10.1046/j.1365-8711.2000.03715.x, Scoccimarro_2001,2002ApJ...575..587B, 2003ApJ...593....1B, Zheng_2005}). The HOD framework circumvents the messy physics of baryons by establishing an empirical relation between galaxies and haloes that can be constrained from measurements of galaxy clustering. Further, HODs provide a computationally inexpensive method of ``painting" galaxies onto $N$-body simulations compared to the ``full physics" approach of hydrodynamic simulations \citep{10.1093/mnras/stab2564}. This becomes especially relevant when forecasting the large cosmological volumes of current and next-generation galaxy surveys including Dark Energy Spectroscopic Instrument (DESI; \citealt{desicollaboration2016desiexperimentisciencetargeting}), \textit{Euclid} \citep{2025A&A...697A...1E}, Vera C. Rubin Observatory (\textit{Rubin}; \citealt{Ivezic_2019}), and the Nancy Grace Roman Space Telescope (\textit{Roman}; \citealt{akeson2019widefieldinfraredsurvey}).

In the standard HOD framework, galaxies are sub-categorized into central and satellite populations based on their spatial locations within haloes. A central galaxy is defined as the one residing at the center of mass of the halo, while the satellite galaxies trace the outer gravitational potential. Several parametric HOD models have been proposed in the literature, many of which assume a smoothed step-function for the central HOD and a power law for the satellite HOD (e.g.\ \citealt{2002ApJ...575..587B, Zheng_2005, Zheng_2007}). The five-parameter model proposed in \citet{Zheng_2005} has demonstrated considerable success in reproducing the occupation statistics of galaxies selected above a minimum stellar mass ($M_{\star}$) in hydrodynamic and semi-analytic simulations. On the contrary, galaxy samples selected on the basis of minimum star formation rate (SFR) exhibit distinctive features that are not well-described by the \citet{Zheng_2005} parameterization. The central HODs of SFR-selected galaxy samples are best described by a Gaussian at low masses and a constant (usually) fractional occupation at higher masses, while the satellite HOD follows a smoothed step function at low masses and a power law scaling at higher masses \citep{Geach2012}. An in-depth comparison of these two parameterizations is conducted in Appendix~\ref{sec:appendix_parametric_HOD}.

The parametric approach to HOD modelling provides a straightforward path to empirically constraining the HOD from measurements of the 2PCF. This involves assuming a functional form for the HOD to construct a forward model of the 2PCF. The best-fit parameters of the HOD are then determined by iteratively exploring the HOD parameter space using a sampling algorithm (e.g.,\ Markov Chain Monte Carlo; MCMC), forward modelling each set of HOD parameters into a ``model" 2PCF, and comparing to the observed 2PCF. This method of HOD estimation has been applied extensively in the literature to derive constraints on the HODs of different galaxy samples over a range of redshifts (e.g.\ \citealt{2007ApJ...667..760Z, White_2007, White_2011, Geach2012, 2012A&A...542A...5C, 10.1093/mnras/stv1966, 2015A&A...583A.128D, 2017MNRAS.469.2913C}).

While the parametric approach to HOD modelling allows for an empirical determination of the HOD from measurements of the galaxy 2PCF, degeneracies that arise amongst the HOD model parameters can lead to an underestimation of the HOD uncertainties if the full covariance matrix is not carefully propagated. Further, the choice of parameterization may bias the form of the HOD. This problem hints at a deeper issue of parametric rigidity and implies that existing parameterizations cannot represent the complete space of physically plausible galaxy samples. This arises in part because the shape of the HOD is highly dependent on the physical properties of the selected galaxy sample. This becomes especially relevant when selecting galaxies in a multidimensional physical property space (e.g.\ a simultaneous cut in $M_{\star}$ \textit{and} SFR), as can be necessary to study specific types of galaxies. For example, Ly$\alpha$ Emitters (LAEs; \citealt{2020ARA&A..58..617O}) are selected using Ly$\alpha$ equivalent width and SFR thresholds and H$\alpha$ emitters are often selected using slitless spectroscopy \citep{10.1111/j.1365-2966.2010.16364.x}, which is indirectly influenced by thresholds in $M_{\star}$ and SFR. Performing a traditional parametric HOD inference on these galaxy samples can bias the form of the inferred HOD. 

In this paper, we propose a non-parametric approach to HOD modelling. We do not assume an explicit parametric form for the HOD. Instead, we measure HODs as a function of the physical galaxy property limits that are used to select simulated galaxy samples from cosmological simulations. We then use a Bayesian inference framework to derive constraints on the HOD. In this initial proof-of-principle work, we make the optimistic assumption that the simulated galaxy samples match those that could be observed in the real universe. Because we are no longer restricted to any particular functional form for the HOD, this approach has the potential to infer unbiased, accurate constraints on the HOD of an arbitrary galaxy sample. However, this non-parametric HOD approach is complicated by the significant computational burden involved in computing the galaxy 2PCF,  which has complexity $\propto O(N \log N)$ where $N$ is the number of galaxies in the sample for typical state-of-the-art algorithms, at each point in physical galaxy property space. For this reason, we train a machine learning-based algorithm to predict (i.e.,\ emulate) the 2PCF and HOD as a function of the physical galaxy property limits that are used to select the galaxy sample. Incorporating a trained emulator of the 2PCF and HOD into a MCMC, we can dramatically speed-up each likelihood evaluation, and ensure the computational feasibility of the non-parametric HOD approach. Although machine learning-based approaches to HOD modelling have been explored in the literature (e.g.\ \citealt{Zhai_2019, wadekar2020modelingassemblybiasmachine, delgado2022modeling, dumerchat2023galaxyclusteringmultiscaleemulation, Storey-Fisher_2024, jana2024constraininggalaxyhaloconnectionusing}), all of these analyses adopt an underlying HOD parameterization. Our approach differs from these in that we do not require any assumption on the shape of the HOD \textit{a priori}. 

An alternative empirically motivated approach to studying the galaxy-halo connection is subhalo abundance matching (SHAM; \citealt{2004ApJ...609...35K, 2004MNRAS.353..189V}). The traditional SHAM model employs a similar wisdom to our proposed non-parametric HOD model in that it does not assume an explicit functional form for the central and satellite HOD. SHAM instead adopts an injective and monotonic relation between galaxy properties (often $M_{\star}$) and subhalo properties (e.g., the subhalo circular velocity; \citealt{2006ApJ...647..201C, 2014MNRAS.443.3044Z}). Associating host haloes with central galaxies and subhaloes with satellite galaxies naturally determines the halo occupation statistics of a given threshold sample without needing to model baryonic physics (an analogous feature of HOD models). While traditional SHAM modelling can efficiently predict galaxy clustering, it requires careful extension to account for correlations between stellar mass and any other galaxy properties (e.g., \citealt{Hearin_2013, Chaves_Montero_2016}).

In what follows, we will demonstrate how the HOD can be inferred from measurements of the galaxy 2PCF without assuming a parametric form for the HOD. We recognize that our method is not a \textit{fully} non-parametric HOD model in terms of being able to fit \textit{absolutely} any HOD shape, rather it uses two physical properties as parameters instead of HOD parameters. In Section \ref{sec:2PCF} we outline how the 2PCF is estimated from (simulated) galaxy survey data. In Section \ref{sec:HODs} we introduce the standard HOD paradigm, and derive an expression for the galaxy 2PCF in terms of the HOD. In Section \ref{sec:sims} we describe the suite of cosmological simulations used to train, validate, and test our emulator and introduce a unique approach to efficiently estimating the galaxy 2PCF. In Section \ref{sec:emu_and_inference} we detail the emulation algorithm, training procedure, and introduce the emulator-assisted inference framework used to derive constraints on the HOD from a measurement of the galaxy 2PCF. We validate the HOD recovery performance over a collection of mock observations drawn from the Santa Cruz semi-analytic model-based simulation used to train the emulator in Section \ref{sec:sc_sam_validation}. In Section \ref{sec:tng_results} we test the HOD recovery performance using a set of mock observations drawn from the TNG100-1 simulation. In Section \ref{sec:caveat_discussion} we discuss the methodological limitations and caveats of our non-parametric HOD framework and the direction of future work. In Section \ref{sec:conclusion} we summarize the implications of our results.

\section{Background}\label{sec:background}

\subsection{Galaxy two-point correlation function}\label{sec:2PCF}

The spatial clustering of galaxies is often characterized using the isotropic real space galaxy 2PCF $\xi(r)$. $\xi(r)$ quantifies the excess-over-random probability of finding two galaxies within volume elements $dV_1$ and $dV_2$ at separation $r$ 
\begin{equation}
dP_{12}(r) = \bar{n}^2[1 + \xi(r)] dV_1 dV_2
\end{equation}
where $\bar{n}$ is the mean number density of the galaxy sample \citep{Bernardeau_2002}. In principle, $\xi(r)$ is only measurable when the line-of-sight distance to each galaxy is available. This is true for galaxy redshift surveys, where individual galaxies are labelled by their sky coordinates ($\alpha_i, \delta_i, z_i$). Here, $\alpha_i$ is the right ascension, $\delta_i$ is the declination, and $z_i$ is the redshift. In this work, we treat the redshift $z_i$ as purely cosmological in origin (i.e.,\ the apparent Doppler shift arises solely from the expansion of the Universe), and reserve consideration of peculiar velocity-induced redshift space distortions for future work. To compute each separation $r$, each coordinate ($\alpha_i, \delta_i, z_i$) is transformed into Cartesian form ($X_i, Y_i, Z_i$) and the Euclidean separation between galaxy pairs is calculated. This involves the intermediate step of measuring the comoving distance $\chi(z)$ to each galaxy,
\begin{equation}
    \chi(z)  = c \int_0^z \frac{dz}{H(z)}
    \label{eq:comoving_dist}
\end{equation}
where $c$ is the speed of light and $H(z)$ is the Hubble parameter \citep{hogg2000distancemeasurescosmology}. In this work, we adopt a flat $\Lambda$CDM Universe with the cosmological parameter values reported by the Planck Collaboration in 2015 \citep{Planck2016}. 

In surveys where the line-of-sight distance to each galaxy is not measured (i.e. photometric galaxy surveys), the galaxy catalogue consists only of coordinates ($\alpha_i, \delta_i$). This results in a measurement of the angular two-point correlation function $w(\theta)$, related to $\xi(r)$ via a line-of-sight projection integral. $w(\theta)$ can be defined in a similar manner to $\xi(r)$, representing the excess-over-random probability of finding two galaxies in the solid angle elements $d \Omega_1$ and $d \Omega_2$ with a mean surface number density $\bar{\nu}$ and at an angular separation $\theta$ \citep{1980lssu.book.....P}. The angular separation $\theta$ (in degrees) is related to the real-space separation $r$ as
\begin{equation}
    \theta = \frac{180}{\pi} \left[2 \arcsin\left(\frac{r}{2\chi(z)}\right)\right].
\end{equation} 
In this work, we use $w(\theta)$ as the basis for both the emulator discussed in Section \ref{sec:GP_emu} and the observable space to which we fit. This is because $w(\theta)$ is the nominal observable in photometric galaxy clustering studies over which we plan to validate our methodology in future work.

In practice, $w(\theta)$ is estimated from a data catalogue ($D$) by counting pairs of galaxies as a function of separation distance and comparing with the distribution expected from a random catalogue ($R$) that covers the same geometric footprint as the survey. Various estimators of the 2PCF have been presented in the literature (e.g.\ \citealt{1983ApJ...267..465D, 1993ApJ...417...19H, Landy_Szalay_1993}), the simplest being the Naive estimator originally proposed by \citet{1973ApJ...185..757H};
\begin{equation}
w(\theta) = \frac{DD(\theta)}{RR(\theta)}\frac{N_R (N_R - 1)}{N_D(N_D - 1)}- 1
\label{eq:Naive}
\end{equation}
where $DD(\theta)$ and $RR(\theta)$ are the number of data-data pair counts and random-random pair counts with pair separations in the range $\theta \pm \Delta \theta / 2$, $N_D$ is the total number of galaxies in the data catalogue, and $N_R$ is the total number of points in the random catalogue (in this work, we specify $N_R = 10 N_D$).

Because $N_D$ and $N_R$ are often large (i.e. $ > 10^5$ for large scale galaxy surveys) the 2PCF is generally an expensive statistic to compute. As a result, various algorithms have been employed to speed up the computation of the 2PCF by making approximations on large scales. In this work, we use the Python package \texttt{treecorr}\footnote{\url{https://github.com/rmjarvis/TreeCorr}, v5.0} \citep{2004MNRAS.352..338J}
to compute the $DD(\theta)$ and $RR(\theta)$ histograms. \texttt{treecorr} speeds up the computation of histograms by implementing a ball tree-based routine that partitions data points into smaller subsets to achieve an overall complexity scaling of $O(N\log N)$, where $N$ is the number of galaxies. While this represents a considerable improvement, estimating 2PCFs for a large number of galaxy samples, as is necessary for the inference we will conduct, remains a computationally demanding task. This motivates the use of the emulator that is introduced in Section \ref{sec:emu_and_inference}.

\subsection{Halo Occupation Distributions} \label{sec:HODs}

The standard HOD framework predicts the average number of galaxies (with a specific set of properties) that reside within a dark matter halo using the conditional probability distribution $P(N | M_{{\text{h}}})$ that a halo of virial mass $M_{ {\text{h}}}$ contains $N$ galaxies \citep{2002ApJ...575..587B}. The mean number of galaxies occupying a halo as a function of virial mass (i.e. the first moment of $P(N | M_{{\text{h}}})$) is then
\begin{equation}
\langle N (M_{\text{h}}) \rangle = \sum_{N=1}^{\infty} N P(N | M_{\text{h}}).
\label{eq:HOD}
\end{equation}
A standard assumption of HOD framework stipulates that all haloes with an occupation number $N>0$ contain one central galaxy and $N-1$ satellite galaxies. As a result, the mean occupation number of galaxies can be decomposed into mean central and satellite occupations
\begin{equation}
\langle N(M_{\text{h}}) \rangle = \langle N_{\text{cen}}(M_{\text{h}}) \rangle + \langle N_{\text{sat}} (M_{\text{h}}) \rangle
\end{equation}
where $\langle N_{\text{cen}}(M_{\text{h}}) \rangle$ and $\langle N_{\text{sat}}(M_{\text{h}}) \rangle$ will herein be referred to as the central and satellite HODs, respectively. 

\subsection{Galaxy Clustering and HODs}\label{sec:halo_mod}

The halo model provides a convenient analytic formalism to describe the clustering of galaxies, operating under the principle that all matter is partitioned over haloes that have well-understood properties \citep{2004ApJ...609...35K}. This enables the 2PCF to be specified entirely by 5 components:
\begin{enumerate}
    \item the halo mass function $\frac{dn_{\text{h}}}{dM_{\text{h}}}$ defined such that $\frac{dn_{\text{h}}}{dM_{\text{h}}}dM_{\text{h}}$ is the number of haloes in the mass range $[M_{\text{h}}, M_{\text{h}}+dM_{\text{h}}]$ per unit comoving volume, 
    \item the linear halo bias function $b_{\text{h}}(M_{\text{h}})$,
    \item the isotropic normalized, radial number density profile of satellite galaxies in haloes $u_{\text{s}}(r|M_{\text{h}})$,
    \item the linear 2PCF of dark matter $\xi^\text{lin}_{\text{DM}}(r)$, and
    \item the conditional probability distribution $P(N | M_{\text{h}})$ provided by the HOD formalism.
\end{enumerate}
Following a similar derivation to \citet{COORAY_2002}--and operating in the $\xi(r)$ basis for consistency with the literature--the dependence of the 2PCF on each of the five components listed above can be derived analytically by considering the galaxy over-density field
\begin{equation}
\delta_{\text{g}}(\boldsymbol{x}) = \frac{n_{\text{g}}(\boldsymbol{x})}{\bar{n}_{\text{g}}} - 1
\end{equation}
where $n_{\text{g}}(\boldsymbol{x})$ is the number density of galaxies in real-space and $\bar{n}_{\text{g}} $ is its mean (i.e.,\ $\langle n_{\text{g}}(\boldsymbol{x}) \rangle = \bar{n}_{\text{g}} $). 
The autocorrelation of this field is the galaxy 2PCF
\begin{equation}
\xi(r) = \frac{1}{\bar{n}^2_{\text{g}} } \langle n_{\text{g}}(\boldsymbol{x})n_{\text{g}}(\boldsymbol{x}+\boldsymbol{r}) \rangle - 1.
\label{eq:crosscorr}
\end{equation}
Expressing $n_{\text{g}}(\boldsymbol{x})$ as a summation over the central and satellite galaxy contributions of the $i$th halo with mass $M_{{\text{h}}, i}$ and a center of mass located at $\boldsymbol{x}_i$
\begin{equation}
n_{\text{g}}(\boldsymbol{x}) = \sum_i N_{\mathrm{cen},i} \delta^3 (\boldsymbol{x}-\boldsymbol{x}_i) + N_{\mathrm{sat},i} u_{\text{s}}(|\boldsymbol{x}-\boldsymbol{x}_i| | M_{{\text{h}}, i})
\end{equation}
Equation~\eqref{eq:crosscorr} becomes
\begin{equation} 
\begin{aligned} 
& \xi(r) = \frac{1}{\bar{n}^2_{\text{g}}} \sum_{i,j} \left[ \langle N_{\mathrm{cen},i} N_{\mathrm{cen},j} \delta^3 (\boldsymbol{x}-\boldsymbol{x}_i) \delta^3 (\boldsymbol{x}+\boldsymbol{r}-\boldsymbol{x}_j)\rangle \, +\right. \\ &\left. \langle N_{\mathrm{cen},i} N_{\mathrm{sat},j} \delta^3 (\boldsymbol{x}-\boldsymbol{x}_i) u_{\text{s}}(|\boldsymbol{x}+\boldsymbol{r}-\boldsymbol{x}_j| | M_{{\text{h}}, j})\rangle \, + \right. \\ &\left. \langle N_{\mathrm{sat},i} N_{\mathrm{cen},j} u_{\text{s}}(|\boldsymbol{x}-\boldsymbol{x}_i| | M_{{\text{h}}, i}) \delta^3 (\boldsymbol{x}+\boldsymbol{r}-\boldsymbol{x}_j) \rangle \, + \right. \\ &\left. \langle N_{\mathrm{sat},i} N_{\mathrm{sat},j} u_{\text{s}}(|\boldsymbol{x}-\boldsymbol{x}_i| | M_{{\text{h}}, i})u_{\text{s}}(|\boldsymbol{x}+\boldsymbol{r}-\boldsymbol{x}_j| | M_{{\text{h}}, j}) \rangle \right] - 1  \label{eq:full_2PCF_eq_discrete}\end{aligned}
\end{equation}
Following convention, we split the summation into $i=j$ (one-halo) and $i \neq j$ (two-halo) terms to separate the contribution from pairs of galaxies located in the same halo from those located in separate haloes, such that
\begin{equation}
    \xi(r) =  \xi^{\text{1h}}(r) + \xi^{\text{2h}}(r).
\end{equation}
The one-halo term of Equation \eqref{eq:full_2PCF_eq_discrete} then becomes
\begin{equation}
\begin{aligned}
 \xi^{1\mathrm{h}}(r) = &\frac{1}{\bar{n}^2_{\text{g}}} \sum_{i} \left[\langle N_{\mathrm{cen},i}^2 \delta^3 (\boldsymbol{x}-\boldsymbol{x}_i) \delta^3 (\boldsymbol{x}+\boldsymbol{r}-\boldsymbol{x}_i)\rangle \, +\right. \\ &\left. \langle N_{\mathrm{cen},i} N_{\mathrm{sat},i} \delta^3 (\boldsymbol{x}-\boldsymbol{x}_i) u_{\text{s}}(|\boldsymbol{x}+\boldsymbol{r}-\boldsymbol{x}_i| | M_{{\text{h}}, i})\rangle \, + \right. \\ &\left. \langle N_{\mathrm{sat},i} N_{\mathrm{cen},i} u_{\text{s}}(|\boldsymbol{x}-\boldsymbol{x}_i| | M_{{\text{h}}, i}) \delta^3 (\boldsymbol{x}+\boldsymbol{r}-\boldsymbol{x}_i) \rangle \, + \right. \\ &\left. \langle N_{\mathrm{sat},i}^2 u_{\text{s}}(|\boldsymbol{x}-\boldsymbol{x}_i| | M_{{\text{h}}, i})u_{\text{s}}(|\boldsymbol{x}+\boldsymbol{r}-\boldsymbol{x}_i| | M_{{\text{h}}, i}) \rangle \right]  
\end{aligned}
\end{equation}
where we have implicitly shifted the ``-1" in Equation \eqref{eq:full_2PCF_eq_discrete} to the two-halo term. Expressing the ensemble average and summation as a double integral over mass and space, the one-halo term then becomes
\begin{equation}
\begin{aligned}
\xi^{1\mathrm{h}}(r) = \frac{1}{\bar{n}^2_{\mathrm{g}}} &\Bigg[ 
 2 \int dM_{\mathrm{h}} \frac{dn_{\mathrm{h}}}{dM_{\mathrm{h}}}  
   \langle N_{\mathrm{cen}}(M_{\mathrm{h}}) \rangle 
   \langle N_{\mathrm{sat}}(M_{\mathrm{h}}) \rangle  
   u_{\mathrm{s}}(r \,|\, M_{\mathrm{h}}) \\ 
& + \int dM_{\mathrm{h}} \frac{dn_{\mathrm{h}}}{dM_{\mathrm{h}}}  
   \langle N_{\mathrm{sat}}(M_{\mathrm{h}})(N_{\mathrm{sat}}(M_{\mathrm{h}})-1) \rangle 
   \\ 
& \times \int d\boldsymbol{y} \, 
   u_{\mathrm{s}}(|\boldsymbol{x}-\boldsymbol{y}| \,|\, M_{\mathrm{h}}) \,
   u_{\mathrm{s}}(|\boldsymbol{x}+\boldsymbol{r}-\boldsymbol{y}| \,|\, M_{\mathrm{h}})
\Bigg]
\end{aligned}\label{eq:1h_full_term}
\end{equation}
where we have assumed that the occupation statistics of central and satellite galaxies are independent to separate $\langle N_{\mathrm{cen}} N_{\mathrm{sat}}\rangle = \langle N_{\mathrm{cen}} \rangle \langle N_{\mathrm{sat}}\rangle$ and ignored the contribution from central-central self-pairs, which introduce artificial power at $r=0$ via $\delta^3(r)$ \citep{Asgari_2023}. We can further simplify the second term of Equation \eqref{eq:1h_full_term} if we assume the satellite occupation to follow Poisson statistics
\begin{equation}
\begin{aligned}
\xi^{1\mathrm{h}}(r)  = \frac{1}{\bar{n}^2_{\mathrm{g}}} & \Bigg[ 
 2 \int dM_{\mathrm{h}} \frac{dn_{\mathrm{h}}}{dM_{\mathrm{h}}}  
   \langle N_{\mathrm{cen}}(M_{\mathrm{h}}) \rangle 
   \langle N_{\mathrm{sat}}(M_{\mathrm{h}}) \rangle  
   u_{\mathrm{s}}(r | M_{\mathrm{h}}) \\ 
& + \int dM_{\mathrm{h}} \frac{dn_{\mathrm{h}}}{dM_{\mathrm{h}}}  
   \langle N_{\mathrm{sat}}(M_{\mathrm{h}})\rangle^2 (u_{\mathrm{s}} \ast
   u_{\mathrm{s}})(r | M_{\mathrm{h}})
\Bigg]
\end{aligned}\label{eq:1h_full_term_final}
\end{equation}
where $(u_{\mathrm{s}} \ast u_{\mathrm{s}})(r|M_{\mathrm{h}})$ is the auto-convolution of the radial number density profile of satellites.
The two-halo term can be obtained similarly by considering the $i \neq j$ case of Equation \eqref{eq:full_2PCF_eq_discrete}, although with two additional considerations. Because dark matter haloes are themselves clustered, the halo-halo correlation function $\xi_{\text{hh}}(r | M_{\text{h}, 1}, M_{\text{h}, 2})$ needs to be accounted for. Typically, haloes are expected to follow the clustering of the dark matter density field, albeit with a mass-dependent enhancement called the linear halo bias function $b_{\text{h}}(M_{\text{h}})$ \citep{1996MNRAS.282.1096M, 1999MNRAS.308..119S}. Under this approximation $\xi_{\text{hh}}(r | M_{\text{h}, 1}, M_{\text{h}, 2}) \approx b_{\text{h}}(M_{\text{h}, 1}) b_{\text{h}}(M_{\text{h}, 2}) \xi^\text{lin}_{\text{DM}}(r)$. A further simplification can be made by assuming $u_{\text{s}}(r|M)$  to be sharply-peaked relative to the separations considered in the two-halo regime such that $u_{\mathrm{s}}(\boldsymbol{x}-\boldsymbol{y}|M_{\mathrm{h}}) \approx \delta^3(\boldsymbol{x}-\boldsymbol{y})$, yielding
\begin{equation}
\begin{aligned}
 \xi^{\text{2h}}&(r) \approx \xi^{\text{lin}}_{\text{DM}}(r)
 \left[\frac{1}{\bar{n}_{\text{g}}} \int dM_{\text{h}} \frac{dn_{\text{h}}}{dM_{\text{h}}}  \langle N(M_{\text{h}}) \rangle b_{\text{h}}(M_{\text{h}}) \right]^2.
 \label{eq:2halo_term}
\end{aligned}
\end{equation}
The term in brackets is often referred to as the large-scale galaxy bias $b_{\text{g}}$, such that $\xi^{\text{2h}}(r) = b_{\text{g}}^2 \, \xi^{\text{lin}}_{\text{DM}}(r)$ \citep{1984ApJ...284L...9K}. The two-halo term in angular-space can be obtained by directly substituting the angular linear 2PCF of dark matter $w_{\text{DM}}^{\text{lin}}(\theta)$ for $\xi_{\text{DM}}^{\text{lin}}(r)$, yielding $w^{\text{2h}}(\theta) = b_{\text{g}}^2 \, w^{\text{lin}}_{\text{DM}}(\theta)$. 

The halo model derivation of the one-halo term (which dominates on small scales) and two-halo term (which dominates on large scales) emphasizes the dependence of clustering on the underlying halo occupation statistics. More specifically, the one-halo term depends on the second moment of $P(N|M_{\text{h}})$, $ \langle N(N-1) \rangle = \sum_{N=1}^{\infty} N(N-1)P(N|M_{\text{h}})$ and the two-halo term depends on the first moment of $P(N|M_{\text{h}})$, $\langle N (M_{\text{h}}) \rangle$.

\section{Simulations}\label{sec:sims}

\subsection{SC SAM}\label{sec:sc_sam_sims}

To train and validate our emulator we use the suite of simulated lightcones presented in \citet{10.1093/mnras/stac3595}. These simulations include predicted physical galaxy properties and photometry for the deep-field extragalactic surveys that will be carried out by a variety of space- and ground-based observatories, including the James Webb Space Telescope (JWST; \citealt{2023PASP..135f8001G}), \textit{Roman}, \textit{Euclid}, and \textit{Rubin} in the near-future. Each of the five two-deg$^2$ lightcones have a redshift range of $z \sim 0$ to 10. In the initial investigation presented in this paper, we utilize slices of the lightcones spanning $0.7407 < z < 0.7793$, serving the dual purpose of enabling eventual validation against the SDSS Baryon Oscillation Spectroscopic Survey \citep{2011AJ....142...72E} that covers a similar redshift range \textit{and} ensuring the line-of-sight width aligns with the other simulation we consider in Section \ref{sec:TNG_sims}. We additionally validate our methodology over alternative redshift ranges ($0.1 < z < 0.2$ and $1.3 < z < 1.4$), and find that the results of our analysis do not vary significantly. The underlying dark matter haloes in the lightcones are extracted from the dark matter-only $N$-body SMDPL simulation in the MultiDark suite \citep{10.1093/mnras/stw248}, with dark matter halo merger histories generated on-the-fly with the extended Press-Schechter (ePS) formalism \citep[e.g.][]{1999MNRAS.305....1S}. The galaxy populations within these dark matter haloes are simulated with the versatile Santa Cruz semi-analytic model (SC SAM) for galaxy formation (see \citealt{10.1046/j.1365-8711.1999.03032.x, 10.1111/j.1365-2966.2008.13805.x, 10.1111/j.1365-2966.2012.20490.x, 10.1093/mnras/stv1877} for an in-depth description of the various components of this model and see \citealt{2021MNRAS.502.4858S} and \citealt{2022MNRAS.515.5416Y} for details regarding simulated lightcones). To model 
galaxy formation, the SC SAM uses a combination of physical processes that are either described analytically from theory or calibrated from observations and hydrodynamic simulations \citep{2019MNRAS.483.2983Y, 2021MNRAS.508.2706Y}. The SC SAM provides a wide range of predictions for the physical properties of galaxies. This results in an extensive list of galaxy properties available as criteria to select simulated galaxy samples.
The SC SAM has been shown to well-reproduce a wide range of observational constraints, including one-point distribution functions of $M_{\rm UV}$, $M_*$, and SFR and two-point auto-correlation functions from $0\lesssim z \lesssim 7.5$.
In this preliminary analysis, we select galaxy samples using lower limits on stellar mass ($M_{\star}$) and on the instantaneous star formation rate (SFR). Future work can expand the space of physical galaxy properties considered.

For a given simulated galaxy sample, we compute three histograms; (1) the normalized galaxy pair counts within haloes (i.e., the normalized one-halo pair count distribution) as a function of angular separation--$dd^{\text{1h}}(\theta)$, (2) the number density of central galaxies as a function of halo mass--$n_{\text{cen}}(M_{\text{h}})$, and (3) the number density of satellite galaxies as a function of halo mass--$n_{\text{sat}}(M_{\text{h}})$. These histograms were chosen because they can be combined, along with a few additional quantities, to efficiently estimate both the 2PCF and HOD of a galaxy sample. The HODs are computed as
\begin{align}
    \langle N_{\text{cen}}(M_{\text{h}}) \rangle &= \frac{ n_{\text{cen}}(M_{\text{h}})V_{\text{sim}}}{ N_{\text{h}}(M_{\text{h}}) }  \label{eq:2dhist_cen_HOD}, \\
    \langle N_{\text{sat}}(M_{\text{h}}) \rangle &= \frac{ n_{\text{sat}}(M_{\text{h}})V_{\text{sim}}}{ N_{\text{h}}(M_{\text{h}}) } \label{eq:2dhist_sat_HOD}
\end{align}
where $N_{\text{h}}(M_{\text{h}})$ is the halo mass distribution of the simulation and $V_{\text{sim}}$ is the simulation volume in comoving units. To compute the 2PCF, we first consider decomposing $DD(\theta)$ from Equation \eqref{eq:Naive} into one- and two-halo contributions as $DD(\theta) = DD^{\text{1h}}(\theta) + DD^{\text{2h}}(\theta)$, where $DD^{\text{1h}}(\theta)$ is the unnormalized version of $dd^{\mathrm{1h}}(\theta)$ defined above, and $DD^{\text{2h}}(\theta)$ is the pair count distribution of galaxies located in separate haloes. Therefore Equation \eqref{eq:Naive} can be rewritten as
\begin{equation}
    \begin{aligned}
    w(\theta) &= \frac{dd^{\text{1h}}(\theta)}{RR(\theta)}\frac{N_R(N_R-1)}{2} + \left[ \frac{DD^{\text{2h}}(\theta)}{RR(\theta)}\frac{N_R(N_R-1)}{N_D(N_D-1)} - 1 \right] \label{eq:1h_2h_decomp}
    \end{aligned}
\end{equation}
where we identify the term in brackets as the two-halo term $w^{\text{2h}}(\theta)$. Given we have already derived an analytical approximation for $w^{\text{2h}}(\theta)$, we can conveniently avoid having to numerically estimate $DD^{\text{2h}}(\theta)$ through pair counting. Substituting the angular analogue of Equation \eqref{eq:2halo_term} into Equation \eqref{eq:1h_2h_decomp} produces a computationally efficient equation for estimating $w(\theta)$ that can be obtained from the three aforementioned histograms
\begin{equation}
\begin{aligned}
 w(\theta) & \approx 
\frac{dd^{1\text{h}}(\theta)}{RR(\theta)} \frac{N_R(N_R-1)}{2} \,  + \\ & w^{\text{lin}}_{\text{DM}}(\theta) \left[
\frac{1}{\bar{n}_{D}} 
\int dM_{\text{h}} \frac{dn_{\text{h}}}{dM_{\text{h}}}\langle N(M_{\mathrm{h}})\rangle 
b_{\text{h}}(M_{\text{h}})\right]^2.
\end{aligned}
\label{eq:xi_efficient}
\end{equation}
In the above equation, the number density of galaxies in the sample is
\begin{equation}
\bar{n}_{D} =  \sum_i n_{\text{cen}}(M_{\text{h}, i}) + n_{\text{sat}}(M_{\text{h}, i}).
\label{eq:Ngal}
\end{equation}
We use the Core Cosmology Library Python package \texttt{pyccl}\footnote{\url{https://github.com/LSSTDESC/CCL}, v3.2}
 \citep{Chisari_2019} to compute $b_{\text{h}}(M_{\text{h}})$ and adopt the \citet{Tinker_2010} halo bias model. To determine $w_{\text{DM}}^{\text{lin}}(r)$ we use the heuristic detailed in Section \ref{sec:caveat_discussion}. 

\begin{figure*}
\centering
\includegraphics[width=0.90\textwidth]{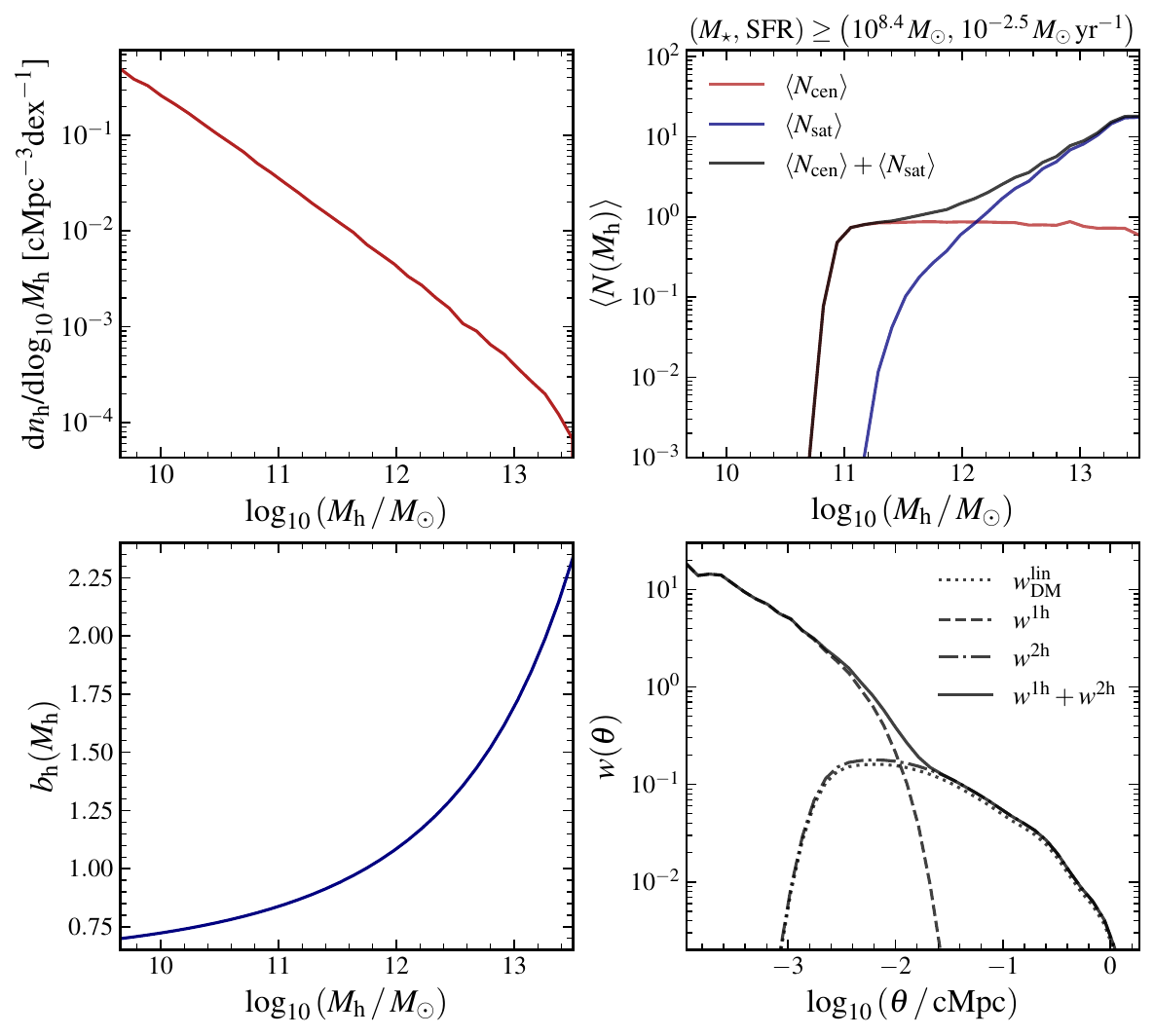}
\caption{Upper left panel: Halo mass function for haloes found in the lightcones at $0.7407<z<0.7793$. Lower left panel: The \cite{Tinker_2010} linear halo bias at $z \sim 0.76$. Upper right panel: The central, satellite, and total HODs for a galaxy sample selected with ($M_{\star}$, SFR) $\geq (10^{8.45} \, M_{\odot}, \, 10^{-2.45} \, M_{\odot} \, \text{yr}^{-1})$ from the five two-deg$^2$ lightcones. Lower right panel: The 2PCF for the same galaxy sample. The one-halo term (dashed) is computed numerically using the first term in Equation \eqref{eq:xi_efficient} while the two-halo term (dashed-dotted) is estimated using the approximation in the second term of Equation \eqref{eq:xi_efficient}. The two-halo term is a rescaling of the linear dark matter two-point correlation function (dotted). The linear dark matter correlation function is truncated on angular scales $\lesssim 10^{-3}$ deg after imposing the scale-dependent additive correction detailed in Section \ref{sec:caveat_discussion}. We emphasize that at small scales the dark matter correlation function represents a highly subdominant contribution (i.e. $\lesssim$ a few percent) to the total galaxy 2PCF.} \label{fig:example_non_parametric_HOD_2PCF}
\end{figure*}

Equation \eqref{eq:xi_efficient} is particularly useful because it naturally isolates the one-halo term from the two-halo term, allowing one to study the behaviour of each component separately. In Figure \ref{fig:example_non_parametric_HOD_2PCF}, we show the form of $\frac{dn_{\text{h}}}{dM_{\text{h}}}$, $b_{\text{h}}(M_{\text{h}})$, and $w_{\text{DM}}^{\text{lin}}(\theta)$ used in this work. We also show the HOD and 2PCF (with its one- and two-halo decomposition) for a galaxy sample with $(\log_{10}(M_{\star}/M_{\odot}),\,  \log_{10}(\mathrm{SFR}/M_{\odot}\mathrm{yr}^{-1})) \geq (8.45, \, -2.45)$. The halo mass function, HOD, and 2PCF curves shown in Figure \ref{fig:example_non_parametric_HOD_2PCF} are the means across the five lightcone realizations considered.

\subsection{IllustrisTNG}\label{sec:TNG_sims}

\begin{figure*}
\centering
\includegraphics[width=0.95\textwidth]{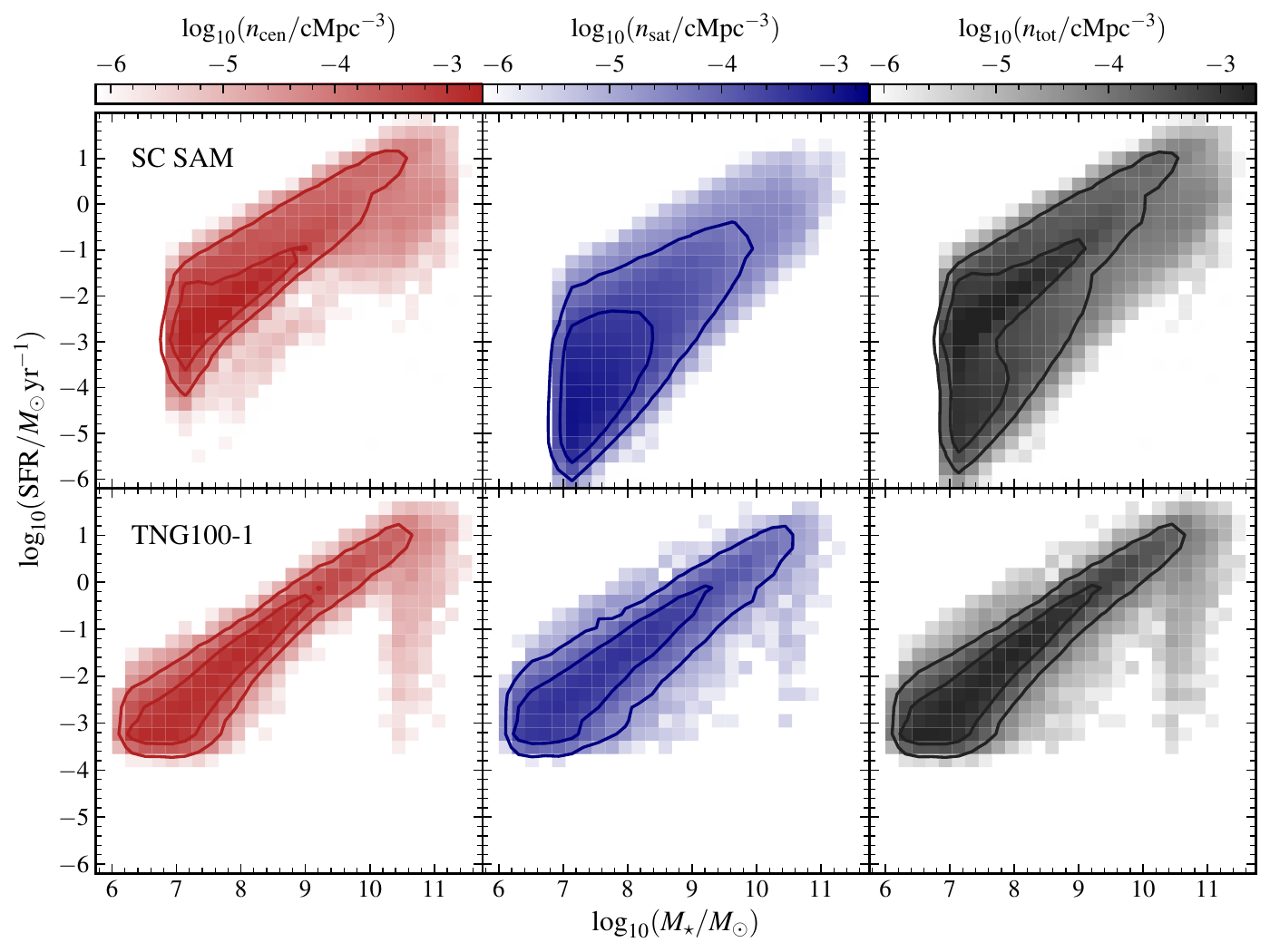}
\caption{SFR-$M_{\star}$ number density distributions of central galaxies (red), satellite galaxies (blue), and their sum (black) in the SC SAM (upper row) and TNG100-1 snapshot (lower row). Contours containing 68\% (inner) and 95\% (outer) of the respective distributions are plotted overtop in each panel. The SC SAM does not resolve galaxies below $\log_{10}(M_{\star}/M_{\odot})=7$, whilst TNG100-1 has an effective minimum resolved SFR of $\log_{10}(\mathrm{SFR}/M_{\odot}\mathrm{yr}^{-1})\approx -3.5$.} \label{fig:mstar_sfr_cen_sat_TNG_SC_comp}
\end{figure*}

To test our methodology, we use the IllustrisTNG project's TNG100-1 simulation \citep{nelson2021illustristngsimulationspublicdata}. TNG100-1 is the main high-resolution IllustrisTNG100 run that utilizes the full TNG physics model. For consistency with the SC SAM lightcones, we select the nearest redshift snapshot from the TNG100-1 simulation at $z=0.76$ which has a cubic volume of $110.7^3$ cMpc$^3$. In TNG100-1, group (i.e., dark matter halo) fields are derived using the friends-of-friends (FoF; \citealt{1985ApJ...292..371D}) algorithm with a linking length of $b=0.2$ \citep{nelson2021illustristngsimulationspublicdata} and subhalo (i.e., galaxy) fields are derived using the SUBFIND algorithm \citep{10.1046/j.1365-8711.2001.04912.x}. To ensure a fair comparison between the SC SAM and TNG100-1, we implement a completeness cut in halo mass, given TNG100-1 has a lower mass resolution and a larger volume than the SC SAM. This restricts the range of halo masses considered in the two simulations to $\log_{10}(M_{\mathrm{h}} / M_{\odot}) \in [9.55, 13.5]$. As alluded to in Section \ref{sec:sc_sam_sims}, we select a $\approx 110.7$ cMpc slice of the SC SAM along the line-of-sight axis that is centered on the TNG100-1 snapshot redshift of $0.76$. This ensures that comparisons between measurements of $w(\theta)$ in the SC SAM and TNG100-1 employ the same amount of projection along the line-of-sight. As detailed in Section \ref{sec:sc_sam_sims} for the SC SAM, we similarly consider TNG100-1 galaxy samples selected above lower bound thresholds in $M_{\star}$ and SFR. A comparison of the two-dimensional $M_{\star}$ and SFR number density distributions of the two simulations is shown in Figure \ref{fig:mstar_sfr_cen_sat_TNG_SC_comp}. Notably, the SC SAM employs a strict threshold at $\log_{10}(M_{\star}/M_{\odot})=7$ below which galaxies are not resolved \citep{10.1093/mnras/stac2139}. Further, TNG100-1 has an effective minimum SFR of $\approx 7 \times 10^{-4}$ $M_{\odot} \mathrm{yr}^{-1}$ imposed by the baryonic mass resolution ($1.4 \times 10^6$ $M_{\odot}$) and the star formation model. Therefore, we impose minimum lower bound thresholds of $\log_{10}(M_{\star}/M_{\odot}) = 7$ and $\log_{10}(\mathrm{SFR}/M_{\odot} \,  \mathrm{yr}^{-1})=-3$ on the galaxy populations considered in this work. Notably, the SC SAM satellite distribution falls off steeply below $\log_{10}(\mathrm{SFR}/M_{\odot}\mathrm{yr}^{-1}) \approx -1.5$ due to quenching (see Section \ref{sec:caveat_discussion} for further discussion). We also note that while the definitions of $M_{\star}$, SFR, and halo mass differ slightly in the two simulations, an effort has been made to compare the most analogous quantities across the SC SAM and TNG100-1, following the discussion in Appendix A of \cite{10.1093/mnras/stac2297}.

\section{Emulation and Inference Framework}\label{sec:emu_and_inference}

\subsection{Gaussian Process-based Emulation}\label{sec:GP_emu}

To learn the mapping between the 2PCF space, the galaxy physical property space, and the HOD space, we use a Gaussian process-based emulator. A Gaussian process (GP) is a collection of random variables, any finite number of which have a joint Gaussian distribution (see \citealt{rasmussen2006gaussian} for an in-depth discussion of GPs). GPs are well-suited to our task given their demonstrated success at efficient, non-linear interpolation across high-dimensional parameter spaces, whilst reporting uncertainties.

\begin{figure*}
\centering
\includegraphics[width=0.90\textwidth]{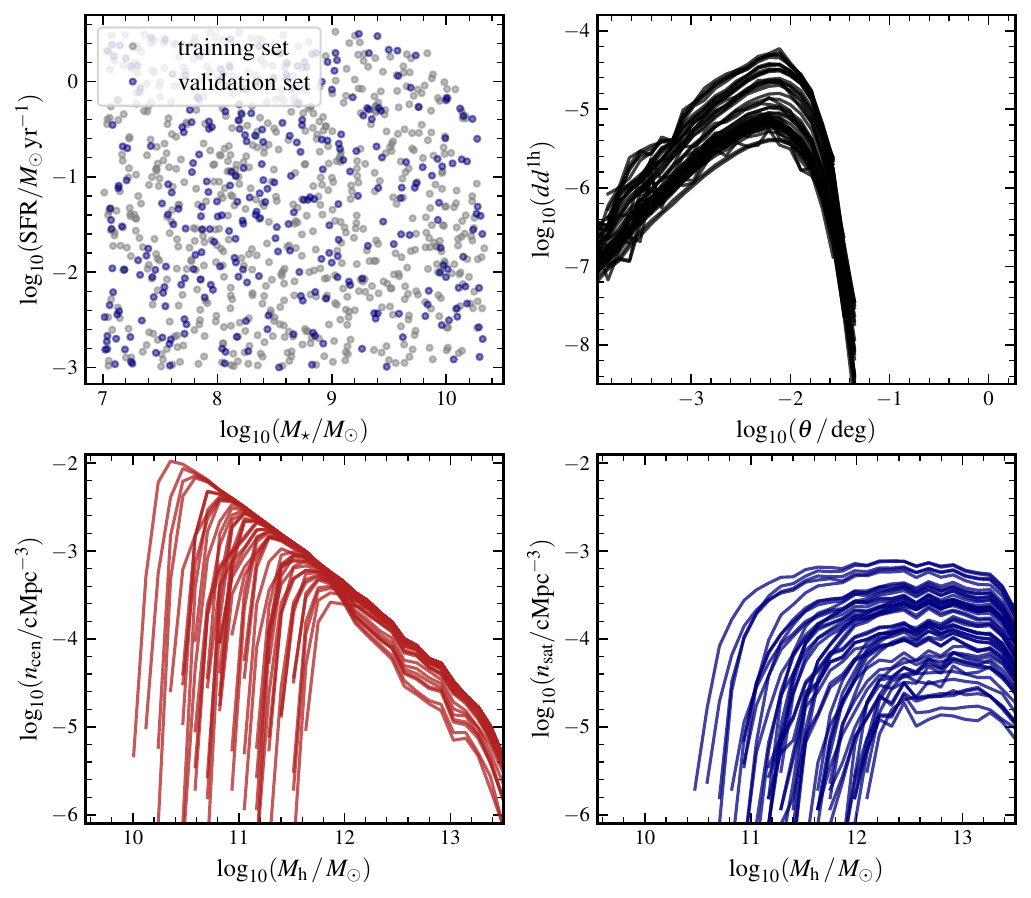}
    \caption{Upper left panel: Training set (grey, 700 samples) and validation set (blue, 300 samples) tuples of physical galaxy property lower bound thresholds. The imposed limit on the minimum galaxy sample size manifests as a restriction on the joint maximum lower bound ($\log_{10}\left(M_{\star}\right), \log_{10}\left(\mathrm{SFR}\right)$) thresholds seen in the upper right hand corner of this panel. Upper right panel: The normalized galaxy pair counts within haloes as a function of separation distance for 40 randomly selected physical galaxy property threshold tuples in the training set. Lower left panel: The number density of central galaxies as a function of halo mass for the same 40 randomly selected physical galaxy property threshold tuples. Lower right panel: The number density of satellite galaxies as a function of halo mass for the same 40 randomly selected physical galaxy property threshold tuples.}
    \label{fig:LHS_sampling}
\end{figure*}

We construct our emulator using the Python package \texttt{GPy}\footnote{\url{https://github.com/SheffieldML/GPy}, v1.13}
\citep{gpy2014}. We initialize an independent GP for each spatial separation bin and halo mass bin of the three histograms introduced above, which are shown in Figure \ref{fig:LHS_sampling}. We construct a training set design matrix $X$ for each GP with a shape $N_{\text{input}} \times  N_{\text{train}}$, where $N_{\text{input}}$ is the number of physical galaxy properties considered (i.e.,\ 2) and $N_{\text{train}}$ is the size of the training set. We additionally construct a training set target vector $\boldsymbol{y}$ with length $N_{\text{train}}$, given that our task is designed to predict a scalar output. Assuming that the targets have additive Gaussian distributed noise, each target $y \in \boldsymbol{y}$ has the form $y = f(\boldsymbol{x}) + \epsilon$ where $\epsilon \sim \mathcal{N}(0, \sigma_n^2)$ and $f$ is the function we aim to learn. We assume $f$ to have zero mean and a covariance described by the kernel function $k$. The prior over $\boldsymbol{y}$ is specified by the covariance matrix $\text{cov}(\boldsymbol{y}) = K(X, X) +\sigma_n^2 I$. Here, $K(X, X)$ is a matrix with shape $N_{\text{train}} \times N_{\text{train}}$, with entries $k(\boldsymbol{x_i}, \boldsymbol{x_j})$ describing how strongly correlated pairs of points $\boldsymbol{x_i}, \boldsymbol{x_j} \in X$ are based on the kernel $k$. In this work, we specify $k$ to be the squared exponential kernel
\begin{equation}
    k(\boldsymbol{x_i}, \boldsymbol{x_j}) = \sigma_k^2 \, \text{exp} \left[- \frac{| \boldsymbol{x_i} - \boldsymbol{x_j}|^2}{2l^2} \right]
\end{equation}
where $\sigma_k$ controls the amplitude of the kernel function, $| \dots |$ is the Euclidean norm, and $l$ is the lengthscale parameter. $\sigma_k$ and $l$ will serve as the hyperparameters that we aim to optimize when training the GP. We similarly construct a validation set design matrix $X_{*}$ with shape $N_{\text{input}} \times N_{\text{val}}$ and a target vector $\boldsymbol{f_*}$ of length $N_{\text{val}}$, where $N_{\text{val}}$ is the size of the validation set. The joint distribution of the training and validation set targets can then be written as
\begin{equation}
    \begin{bmatrix} \boldsymbol{y} \\ \boldsymbol{f_*}
    \end{bmatrix}
    \sim \mathcal{N} \left( \boldsymbol{0}, 
\begin{bmatrix} 
K(X, X) + \sigma_n^2 I & K(X, X_*) \\ 
K(X_*, X) & K(X_*, X_*)
\end{bmatrix} 
\right)
\end{equation}
where $K(X_*, X), \, K(X, X_*),$ and $K(X_*, X_*)$ are matrices defined similarly to $K(X, X)$ and with shapes dependent on the dimensionality of their respective input design matrices. 

The conditional distribution for $\boldsymbol{f_*}$ is then 
\begin{equation}
    \boldsymbol{f_*} | X, \boldsymbol{y}, X_* \sim \mathcal{N}\left(K(X_*, X)[K(X, X)+\sigma_n^2 I]^{-1} \boldsymbol{y},  \text{cov}(\boldsymbol{f_*})\right)
\end{equation}
where 
\begin{equation}
\begin{aligned}
    \text{cov}(\boldsymbol{f_*}) & = K(X_*, X_*) \, - \\ & K(X_*, X) [ K(X, X)+\sigma_n^2 I]^{-1} K(X, X_*).
\end{aligned}
\end{equation}
Training GPs for each bin of $dd^{\text{1h}}(\theta)$, $n_{\text{cen}}(M)$, and $n_{\text{sat}}(M)$ amounts to finding the set of kernel hyperparameters $(\sigma_k, \, l)$ that maximize the log-marginal likelihood
\begin{equation}
    \begin{aligned}
    \ln \left(\boldsymbol{y} | X, (\sigma_k, \, l)\right) = -\frac12 \boldsymbol{y}^{\text{T}} (K(X, X)+\sigma_n^2 I)^{-1} \boldsymbol{y}   \\ - \frac12  \ln \left(\text{det}\left[K(X, X)+\sigma_n^2 I\right] \right) - \frac{N_{\text{train}}}{2} \ln \left(2\pi\right).
    \end{aligned}
\end{equation}
This optimization is performed using the L-BFGS-B algorithm implementation in \texttt{SciPy}\footnote{\url{https://docs.scipy.org/doc/scipy/}, v1.12}  \citep{2020SciPy-NMeth}.

To train and validate the GPs, we generate a near-random sample of 1000 $(\log_{10}\left(M_{\star}\right), \log_{10}\left(\mathrm{SFR}\right))$ tuples using the Latin hypercube sampling algorithm (LHS; \citealt{Mackay_1979}). LHS enables homogeneous and efficient sampling in high-dimensional parameter spaces, whereby each sample lies on its own unique axis-aligned hyperplane. While a simple random sampling algorithm can easily cover the two-dimensional parameter space we consider here, LHS will become especially useful when scaling to higher-dimensional spaces of galaxy properties. In this work, we consider lower bound thresholds spanning the ranges of $\log_{10}\left(M_{\star}/M_{\odot}\right) \in [7.00, 10.33]$ and $\log_{10}\left(\mathrm{SFR}/M_{\odot} \mathrm{yr}^{-1} \right) \in [-3.00, 0.52]$. Tuples that would generate samples of less than 1000 galaxies are excluded, as can be seen in the top right corner of the upper left panel of Figure \ref{fig:LHS_sampling}. This ensures that the galaxy samples we analyze are representative of those typically encountered in real galaxy survey data. We randomly select 700 tuples for the training set and 300 tuples for the validation set (i.e.,\ $N_{\text{train}}=700$, $N_{\text{val}} = 300$). We emphasize that the training and validation sets are generated using samples drawn from the SC SAM (and \textit{not} TNG100-1). The training set is used to obtain the optimal set of model weights for the emulator.

The physical galaxy property tuples corresponding to the training and validation sets are shown in the top left panel of Figure~\ref{fig:LHS_sampling}, while example $dd^{\text{1h}}$, $n_{\text{cen}}$, and $n_{\text{sat}}$ distributions from the training set are shown in the remaining three panels. The emulator's performance evaluated over the validation set is discussed in Section \ref{sec:emu_perf}. We specify $dd^{\text{1h}}$ to span an angular range of $\log_{10}(\theta /\text{deg}) \in [-3.95, 0.27]$ and $n_{\text{cen}}$ and $n_{\text{sat}}$ to span a mass range of $\log_{10}\left(M_{\text{h}}/M_{\odot}\right)\in [9.66, 13.5]$. This minimum angular scale ($\theta \approx 0.40$ arcsec) lies above the minimum measurable separation for current-generation (e.g.\ JWST) space-based imaging, and just below the minimum measurable separation of upcoming ground-based surveys (e.g.\ \textit{Rubin}). We operate in the logarithmic space of both the inputs (physical galaxy property tuples) and the outputs ($dd^{\text{1h}}$, $n_{\text{cen}}$, and $n_{\text{sat}}$ histograms) to ensure numerical stability when training and validating the emulator. For this reason, we quantify the emulator performance in logarithmic space for the three histograms in Figure \ref{fig:emu_performance}.

\subsection{Emulator Performance}\label{sec:emu_perf}

\begin{figure*}
\centering
\includegraphics[width=0.975\textwidth]{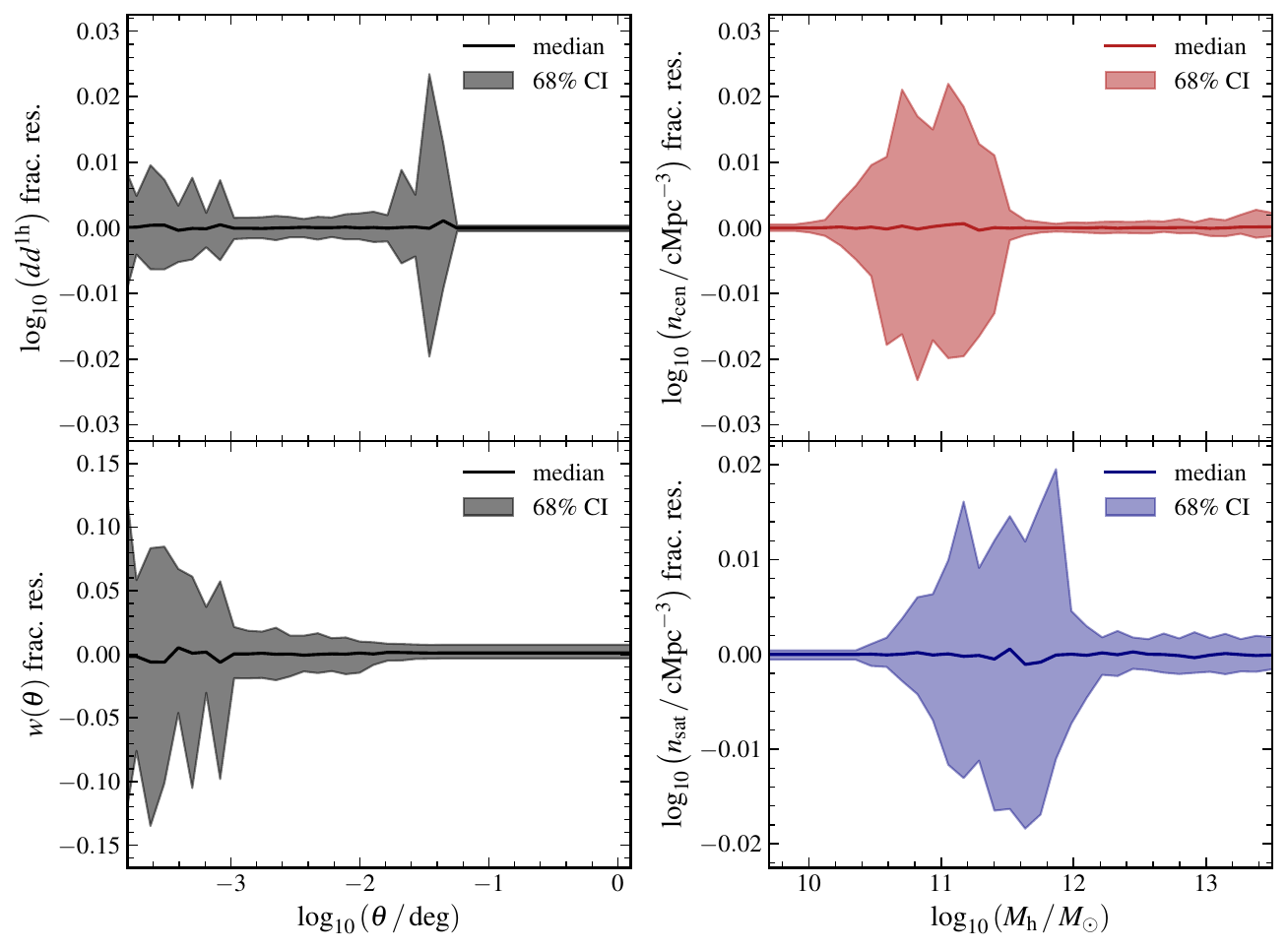}
\caption{Median fractional residuals (solid curves) and 68\% confidence intervals (shaded regions) for $\log_{10}(dd^{1\text{h}})$ (upper left), $w(\theta)$ (bottom left), $\log_{10}(n_{\text{cen}})$ (upper right), and $\log_{10}(n_{\text{sat}})$ (bottom right) computed using the GP-based emulator over the 300 validation set samples. The fractional error reaches a maximum of $\sim 5\%$ across all angular separation and mass bins for the three one-dimensional histograms, indicating the GP has reliably learned the mapping, even with the limited size of the training set. The fractional error in $w(\theta)$ increases from $<3\%$ on intermediate-to-large angular scales to $\approx 5$-$10\%$ on angular scales $\lesssim 10^{-3}$ deg. Note that we have plotted the fractional residuals in $w(\theta)$ and not $\log_{10}(w(\theta))$ as they are more interpretable.} \label{fig:emu_performance}
\end{figure*}

To characterize the predictive performance of the emulator we compute the fractional residual in the $i$th bin of each of the emulated quantities (denoted here as $x$) as, 
\begin{equation}
     \text{(fractional residual)}_{x, i} = \frac{x_{\text{truth},i}-x_{\text{emu},i}}{x_{\text{truth}, i}}
     \label{eq:frac_res}
\end{equation}
where $x_{\text{emu},i}$ is the emulator's prediction and $x_{\text{truth},i}$ is the value in the validation set. The median fractional residuals in each of the emulated quantities $[ \log_{10}(dd^{1\text{h}})$, $\log_{10}(n_{\mathrm{cen}}),$ $\log_{10}(n_{\mathrm{sat}})]$ and their respective 68\% confidence intervals (CI) computed over the 300 validation set samples are shown in Figure \ref{fig:emu_performance}. We also quantify the emulator performance with respect to $w(\theta)$ in Figure \ref{fig:emu_performance}, given it is the key observable. This is done by forward modelling the three histograms into $w(\theta)$ using Equation \eqref{eq:xi_efficient}, and computing the fractional residuals over the same 300 validation set samples. 

The median fractional residuals across all angular separation and halo mass bins do not exceed 5\% and are typically $<$1\% for the three one-dimensional histograms in Figure \ref{fig:emu_performance}. Further, the residuals appear to be symmetrically distributed about 0. Note that the regions where the fractional residuals are $\approx 0$ most often correspond to bins where the distributions are either all 0 or approach an approximately constant value that is easily learned by the GP (e.g.\ large real-space separation bins have predictable, slowly varying behaviour in the two-halo regime, and the occupation statistics of low mass haloes decay to zero). The fractional error in $w(\theta)$ increases from $<3\%$ on intermediate-to-large angular scales to $5$-$10\%$ on smaller angular scales. Satisfied with this level of performance, the emulator can be reliably used as a surrogate model for the 2PCF and HOD within an MCMC framework.

\subsection{Emulator-accelerated MCMC}\label{sec:MCMC}

In order to constrain the HOD for a given 2PCF, we use a Markov Chain Monte Carlo (MCMC). Equipped with a set of sufficiently trained GPs, we have an efficient and reliable emulator of the 2PCF and HOD over the space of $(M_{\star}, \text{SFR})$ thresholds. Incorporating this emulator into the MCMC framework, we can dramatically speed-up each likelihood evaluation and thus reduce the overall time necessary for convergence. We find that the emulator reduces the time for a single likelihood evaluation by a factor of $\gtrsim500$ compared to brute-force computing the 2PCF at each evaluation. 

To implement the MCMC, we use the Python package \texttt{emcee}\footnote{\url{https://github.com/dfm/emcee}, v3.1} \citep{Foreman_Mackey_2013} and adopt the default affine-invariant ensemble sampling algorithm originally introduced in \citet{2010CAMCS...5...65G}. We configure the MCMC with a proposal distribution that includes a combination of the ``stretch move" and ''differential evolution" proposals (see \texttt{emcee} documentation for details). We tune the step sizes for each proposal distribution along with their respective frequencies to obtain a mean acceptance fraction in the range of $0.2$-$0.5$.

As discussed in Section \ref{sec:2PCF}, the observable space in this work is $w(\theta)$. From a Bayesian perspective, the posterior probability distribution over the HOD and the lower bound thresholds $\boldsymbol{G} = (\log_{10}(M_{\star}), \log_{10}(\text{SFR}))$ given a mock observation of the 2PCF $w^{\text{obs}}(\theta)$ is
\begin{equation}
\begin{aligned}
    p&(\langle N(M_{\text{h}}) \rangle, \boldsymbol{G} \, | \, w^{\text{obs}}(\theta)) \propto \\ 
    &p(w^{\text{obs}}(\theta) \, | \, \langle N(M_{\text{h}}) \rangle, \boldsymbol{G})p(\boldsymbol{G})p(\langle N(M_{\text{h}}) \rangle).
\end{aligned}
\end{equation}
We assume a Gaussian likelihood of the form
\begin{equation}
\begin{aligned}
    &p(w^{\text{obs}}(\theta) \, | \, \langle N(M_{\text{h}}) \rangle, \boldsymbol{G}) \propto  \\ &\exp \left[-\frac{1}{2} \left(w^{\text{mod}}(\theta) - w^{\text{obs}}(\theta) \right)^{\text{T}} \mathbf{C}^{-1} \left(w^{\text{mod}}(\theta) - w^{\text{obs}}(\theta) \right) \right]
\end{aligned}
\end{equation}
where $w^{\text{mod}}(\theta)$ is the 2PCF computed at $\boldsymbol{G}$ using the emulator and Equation \eqref{eq:xi_efficient}. The covariance matrix $\mathbf{C} = \mathbf{C}^{\text{obs}}+\mathbf{C}^{\text{mod}}$, with $\mathbf{C}^{\text{mod}}$ the sum of the cosmic variance (here, quantified through the sample variance amongst the five separate lightcone realizations for SC SAM mocks) and the emulator's predicted variance (obtained by propagating the GP's variance predictions on the emulated quantities through Equation \eqref{eq:xi_efficient}). $\mathbf{C}^{\text{obs}}$ is the covariance matrix of the mock observation estimated using the jackknife resampling algorithm with 64 patches in \texttt{treecorr} (see \citealt{10.1111/j.1365-2966.2009.14389.x} for an in-depth discussion of the jackknife method in the context of galaxy 2PCFs). The number of patches was chosen after testing jackknife resampling with 4, 16, 64, and 128 patches and finding that the estimated covariance matrix converged at 64. This covariance matrix serves as a placeholder for the covariance matrix of observational uncertainties when comparing to real survey data. We specify a uniform prior over $\boldsymbol{G}$ with bounds set by the parameter ranges considered in the training set. Given we do not wish to bias the inferred HOD by imposing a restrictive prior on its shape, we specify a Gaussian prior on the galaxy number density in its place. This prior also serves to break degeneracies amongst similar 2PCFs across different regions of parameter space. The logarithm of the posterior the MCMC aims to maximize is then
\begin{equation}
\begin{aligned}
    \ln p(\langle N(M_{\text{h}}) \rangle, \boldsymbol{G} | w^{\text{obs}}(\theta))  \propto \\ -\frac{1}{2} \left(w^{\text{mod}}(\theta) - w^{\text{obs}}(\theta) \right)^{\text{T}} \mathbf{C}^{-1} \left(w^{\text{mod}}(\theta) - w^{\text{obs}}(\theta) \right) \\ - \frac{\left(n_{\text{g, mod}} - n_{\text{g, obs}}\right)^2}{2(\sigma^2_{n_{\text{g, obs}}}+\sigma^2_{n_{\text{g, mod}}})} - \frac12\ln \left(\text{det}\left[\mathbf{C}\right]\right) 
    \\ - \sum_{i=1}^{N_{\text{input}}} \ln \Theta(\boldsymbol{G}_i - \boldsymbol{G}_{i, \text{min}})+ \ln \Theta(\boldsymbol{G}_{i, \text{max}}-\boldsymbol{G}_i). \label{eq:posterior}
\end{aligned}
\end{equation}
Here, $n_{\text{g, mod}}$ is the galaxy number density corresponding to $w^{\text{mod}}$, $n_{\text{g, obs}}$ is the galaxy number density corresponding to $w^{\text{obs}}$, $\sigma^2_{n_{\text{g, mod}}}$ is the variance in $n_{\text{g,mod}}$, 
and $\sigma^2_{n_{\text{g, obs}}}$ is the variance in $n_{\text{g, obs}}$ (taken to be the variance in number density across the five lightcone realizations). The final term enforces the uniform top-hat prior on the physical galaxy properties, where $\boldsymbol{G}_{i, \text{min}}$ and $\boldsymbol{G}_{i, \text{max}}$ denote the minimum and maximum values of the top-hat prior for the $i$th component of $\boldsymbol{G}$, respectively.

We supply this form of the posterior to the \texttt{emcee} affine-invariant ensemble sampler to explore the joint posterior of the $\left(M_{\star}, \text{SFR}\right)$ thresholds. We initialize 8 walkers in a random uniform fashion across the range of $\boldsymbol{G}$ thresholds discussed in Section \ref{sec:GP_emu}, and run the sampler for 1000 steps with 200 steps discarded as burn-in. We additionally discard any walkers with an acceptance fraction $< 0.1$. To determine the HOD that corresponds to the recovered joint posterior over $M_{\star}$-- $\text{SFR}$ space, we randomly draw 1000 samples from the flattened MCMC chains and use the emulator to compute the central, satellite, and total HOD for each sample. For each of these quantities we report the bin-by-bin median and associated 68\% confidence interval. We quantify the recovery performance in HOD space for the validation set in Section \ref{sec:sc_sam_validation} and the test set in Section \ref{sec:tng_results}.

\section{Validation}\label{sec:sc_sam_validation}

To validate the recovery performance in HOD space we compute the fractional and standardized residuals in the recovered HODs using the 300 SC SAM mock observations in the validation set discussed in Section \ref{sec:GP_emu}. We emphasize that when assessing the MCMC recovery performance, the input $w(\theta)$ observations are computed using \texttt{treecorr} (following Equation \eqref{eq:Naive}) and \textit{not} the approximation in Equation \eqref{eq:xi_efficient}. This ensures that our methodology accurately accounts for the differences between empirically-derived measurements of $w(\theta)$ from galaxy survey data and those obtained via forward modelling. The 300 mock observations are divided into three groups, each containing 100 samples; (1) only $M_{\star}$ lower bound thresholds, (2) only SFR lower bound thresholds, and (3) joint ($M_{\star}$, SFR) lower bound thresholds. The values of the lower bound thresholds are generated in accordance with the LHS routine described in Section \ref{sec:GP_emu}, albeit with an additional modification. To obtain the one-dimensional thresholds from the two-dimensional LHS-generated thresholds, we ``zero-out" SFR ($M_{\star}$) to obtain $M_{\star}$-- (SFR--) only thresholds. When running the MCMC we therefore restrict the physical galaxy property space to the singular dimension that corresponds to the one-dimensional threshold. The one-dimensional thresholds provide a useful test of our framework's generalizability but will also serve as a good-faith comparison in Appendix \ref{sec:appendix_parametric_HOD} against the \cite{Zheng_2005} and \cite{Geach2012} parameterizations that are nominally used to model galaxy samples selected in this manner.

To characterize the recovery performance in ($M_{\star}$, SFR) space, we compute both the arithmetic and standardized residuals in the recovered ($M_{\star}$, SFR) tuples using the set of 300 mock observations discussed in Section \ref{sec:GP_emu}.  The arithmetic residuals are computed as $(x_{\text{truth},i} - x_{\text{50}, i})$, where $x_{\text{truth},i}$ is the ground-truth value corresponding to the mock observation and $x_{\text{50}, i}$ is the median of the recovered quantity. The standardized residuals are computed as
\begin{equation}
     \text{(standardized residual)}_{x, i} = \frac{x_{\text{truth},i}-x_{\text{50},i}}{\sigma_{x,i}}
     \label{eq:stand_res}
\end{equation}
where $\sigma_{x,i}$ is the upper or lower bound on the 68\% confidence interval that corresponds to the sign of the residual in the numerator (given the recovered 68\% confidence intervals are generally asymmetric). We note that for one-dimensional thresholds we report only the recovery performance with respect to the acting threshold (e.g., for $M_{\star}-$only thresholds we do not try to recover the lower limit on SFR). 

\begin{figure}
\centering
\includegraphics[width=0.975\columnwidth]{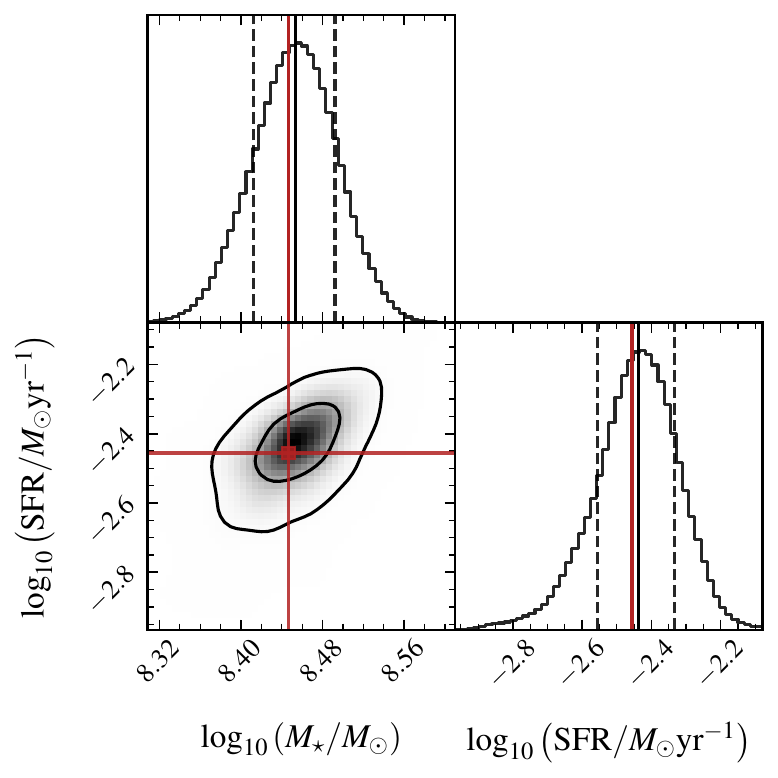}
\caption{Corner plot from a single MCMC run for a SC SAM mock observation with thresholds $(\log_{10}(M_{\star}/M_{\odot})$, $\log_{10}(\mathrm{SFR}/M_{\odot} \mathrm{yr}^{-1})) 
\geq (8.45, -2.45)$ indicated with red solid lines. 
The black lines in the one-dimensional posteriors indicate the 16th (left dashed), 50th (solid), 
and 84th (right dashed) percentiles, with recovered values 
$\log_{10}(M_{\star} / M_{\odot}) = 8.45^{+0.04}_{-0.04}$ and 
$\log_{10}(\mathrm{SFR} / M_{\odot} \mathrm{yr}^{-1}) = -2.44^{+0.10}_{-0.12}$. 
The inner and outer contours in the two-dimensional posterior contain 39.3\% and 86\% of the 
samples, respectively (i.e., one- and two-$\sigma$ contours for a two-dimensional Gaussian distribution). In this example, we are able to recover the ``true" thresholds well within $1 \sigma$ in both $\log_{10}(M_{\star})$ and $\log_{10}(\mathrm{SFR})$ and with high precision.} \label{fig:nonparametric_corner_plot}
\end{figure}

\begin{figure}[htbp]
\centering
\includegraphics[width=\columnwidth]{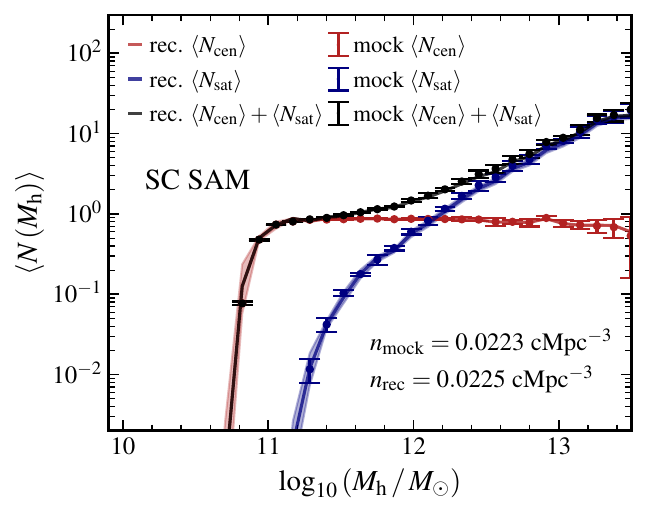}
\caption{Posterior constraints on the recovered median central HOD (red solid), satellite HOD (blue solid), and total HOD (black solid) determined by sampling the MCMC chains in $M_{\star}$-SFR space for the same SC SAM mock observation in Figure \ref{fig:nonparametric_corner_plot}. The respective 68\% confidence intervals on the HODs are indicated as lightly shaded regions. The HODs corresponding to the mock observation are shown as data points, with errors bars representing the sample variance across the five SC SAM lightcone realizations. The significant overlap between the 68\% confidence intervals on the recovered HOD curves and the mock HOD data points indicates that our recovery is both accurate and  precise.} \label{fig:nonparametric_HOD_posterior}
\end{figure}

For a single, randomly-selected MCMC recovery, we show the corner plot over $M_{\star}$--SFR space in Figure \ref{fig:nonparametric_corner_plot} and the corresponding recovered HOD in Figure \ref{fig:nonparametric_HOD_posterior}. Figures \ref{fig:nonparametric_corner_plot} and \ref{fig:nonparametric_HOD_posterior} provide a visual confirmation that our non-parametric HOD approach can simultaneously infer the physical galaxy properties \textit{and} HODs that correspond to a mock observation of the galaxy 2PCF. In Figure \ref{fig:nonparametric_corner_plot}, we are able to identify the maximum of the joint $M_{\star}$--SFR posterior within $1\sigma$ in both $\log_{10}(M_{\star})$ and $\log_{10}(\text{SFR})$ and provide an estimate for its uncertainty. In Figure \ref{fig:nonparametric_HOD_posterior}, we recover both the central and satellite HOD to within $\approx 1 \sigma$ across nearly all of the halo mass bins.

\begin{figure}[htbp]
\centering
\includegraphics[width=\columnwidth]{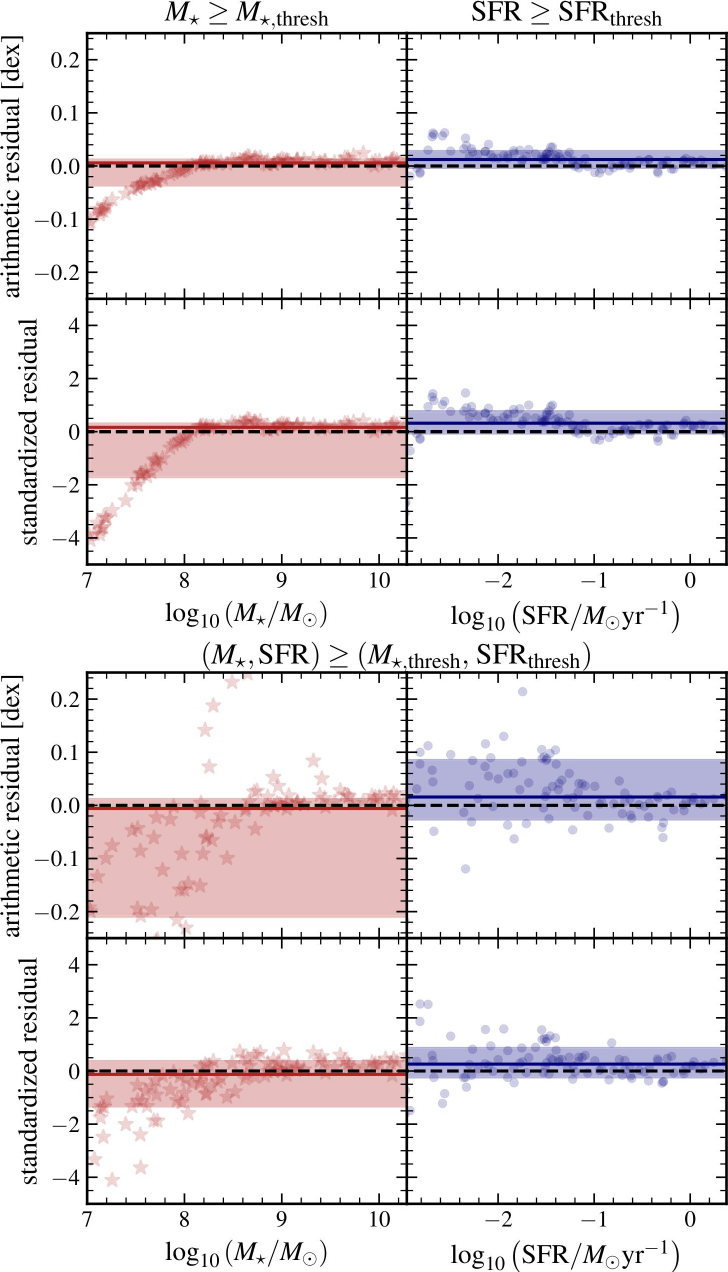}
\caption{Arithmetic and standardized residuals for the true and MCMC-recovered lower bound thresholds on $\log_{10}(M_{\star})$ (left column) and $\log_{10}(\mathrm{SFR})$ (right column) as a function of the ``true" threshold value of the 300 SC SAM mocks for the three types of samples analyzed; only $M_{\star}$ (upper left panels), only $\mathrm{SFR}$ (upper right panels), and joint ($M_{\star}$, SFR) samples (bottom four panels). Residuals from each of the 100 realizations in the three sets are plotted as lightly shaded markers, while the median and 68$\%$ intervals are plotted as solid lines and shaded regions, respectively. We are able to consistently recover both lower bound thresholds within $\approx 0.1$ dex or $\approx 1.5\sigma$.} \label{fig:several_realization_MCMC_parameter_frac_errs}
\end{figure}

The arithmetic and standardized residuals for $\log_{10}(M_{\star})$ and $\log_{10}(\mathrm{SFR})$ computed over all 300 SC SAM mock observations are shown in Figure \ref{fig:several_realization_MCMC_parameter_frac_errs}. In Figure \ref{fig:several_realization_MCMC_parameter_frac_errs}, we find that our approach successfully recovers the lower bound thresholds on all three types of galaxy samples analyzed. For $M_{\star}$--only samples we recover $\log_{10}(M_{\star, \mathrm{thresh}})$ with a median arithmetic residual of $0.006_{-0.043}^{+0.006}$ dex and a median standardized residual of $0.162_{-1.879}^{+0.149}$. For SFR only samples we recover $\log_{10}(\mathrm{SFR}_{\mathrm{thresh}})$ with a median arithmetic residual of $0.012_{-0.016}^{+0.016}$ dex and a median standardized residual of $0.318_{-0.394}^{+0.457}$. For joint ($M_{\star}$, SFR) samples we recover $(\log_{10}(M_{\star, \mathrm{thresh}}), \log_{10}(\mathrm{SFR}_{\mathrm{thresh}}))$ with median arithmetic residuals of $-0.006_{-0.204}^{+0.029}$ dex in $\log_{10}(M_{\star})$ and $0.016_{-0.043}^{+0.070}$ dex in $\log_{10}(\mathrm{SFR})$ and median standardized residuals of $-0.128_{-1.208}^{+0.506}$ in $\log_{10}(M_{\star})$ and $0.259_{-0.502}^{+0.616}$ in $\log_{10}(\mathrm{SFR})$.

\begin{figure*}[htbp]
\centering
\includegraphics[width=\textwidth]{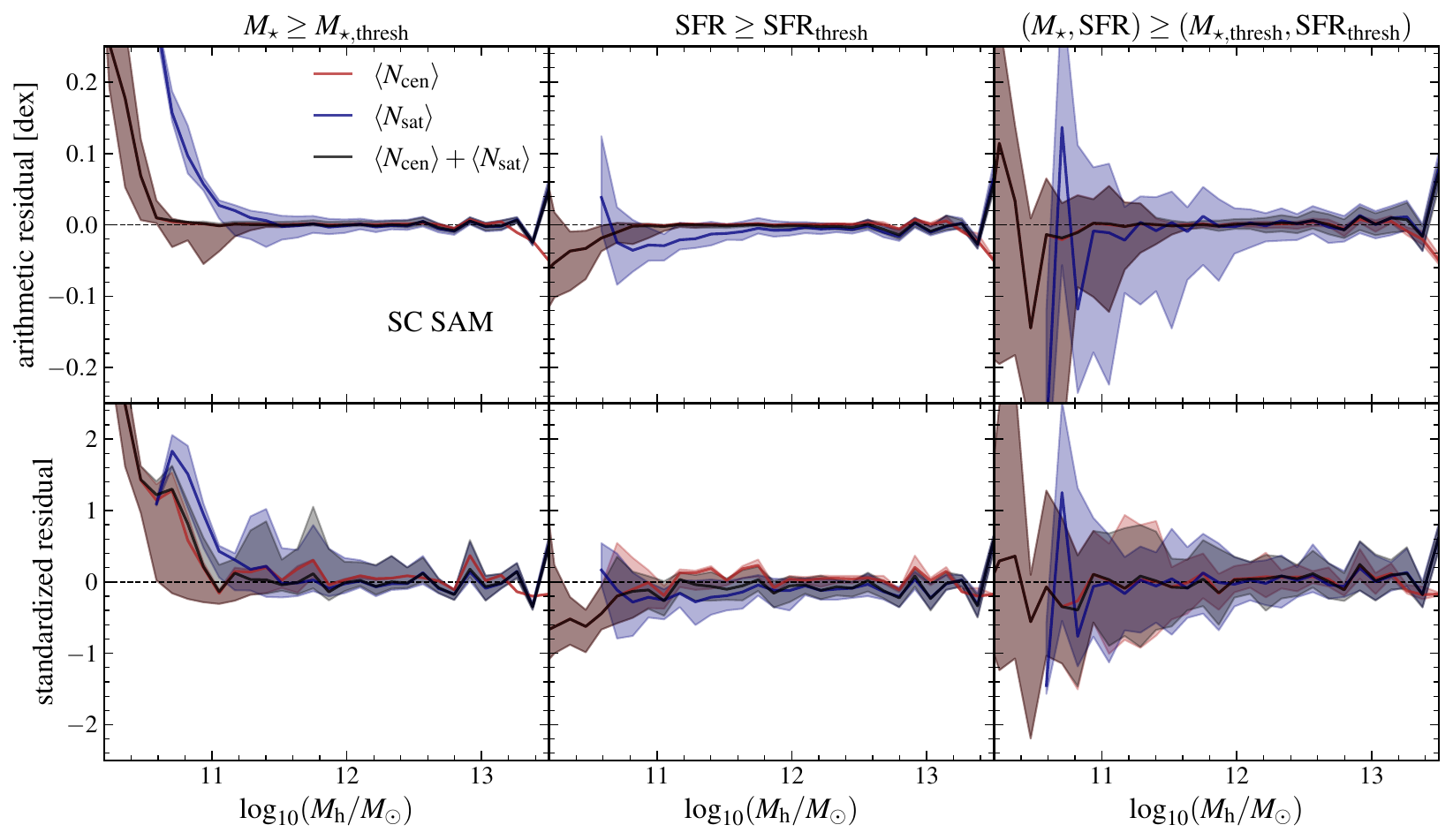}
\caption{Upper row: Arithmetic residuals between the ``true" SC SAM validation set HODs and the recovered HODs obtained using our non-parametric approach. Lower row: Same as the upper row, although now for standardized residuals computed using Equation \eqref{eq:stand_res}. The left column displays residuals for galaxy samples selected with only a $M_{\star}$ lower bound threshold, the middle column displays residuals for samples selected with only a SFR lower bound threshold, and the right column displays residuals for galaxy samples selected above joint ($M_{\star}$, SFR) thresholds. Each of the three columns displays the median arithmetic (upper row) and standardized (lower row) residuals computed across the 100 mock observations that fall into each category as a solid curve, and the corresponding 68\% confidence interval as a shaded region.  We are able to consistently recover the ``true" SC SAM HODs within $\sim 0.1$ dex or $\sim 1\sigma$.} 
\label{fig:several_realization_nonparametric_HOD_posterior_frac_errs}
\end{figure*}

The arithmetic and standardized residuals in the recovered HODs for the 300 SC SAM mock observations are presented in Figure \ref{fig:several_realization_nonparametric_HOD_posterior_frac_errs}. Here, we find that our non-parametric HOD approach consistently recovers the central, satellite, and total HODs with a high degree of accuracy and precision. Accuracy is reflected in the median curves, which exhibit typical arithmetic residuals within $0.1$ dex and standardized residuals within $1\sigma$. Precision is evident from the relatively narrow 68\% confidence intervals in both the arithmetic and standardized residual spaces.

While we have shown that we \textit{can} characterize the recovery performance in physical galaxy property space, we emphasize that the HOD is the primary quantity of interest in this analysis. For this reason, we consider the lower bound thresholds on ($M_{\star}$, SFR) to be nuisance parameters in the context of the TNG100-1 test set, and therefore do not consider their recovery performance in this work.

\section{Results}\label{sec:tng_results}

\begin{figure}[htbp]
\centering
\includegraphics[width=\columnwidth]{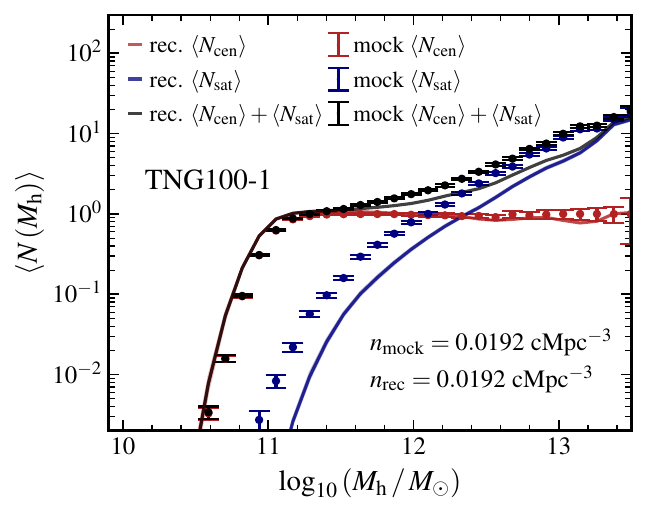}
\caption{Same as Figure \ref{fig:nonparametric_HOD_posterior}, but now for the TNG100-1 mock generated using the same set of $(M_{\star}, \mathrm{SFR})$ lower bound thresholds. The error bars on the mock HOD data points are computed assuming the galaxy counts in each mass bin follow Poisson statistics.} \label{fig:nonparametric_HOD_posterior_TNG100-1}
\end{figure}

\begin{figure*}[htbp]
\centering
\includegraphics[width=\textwidth]{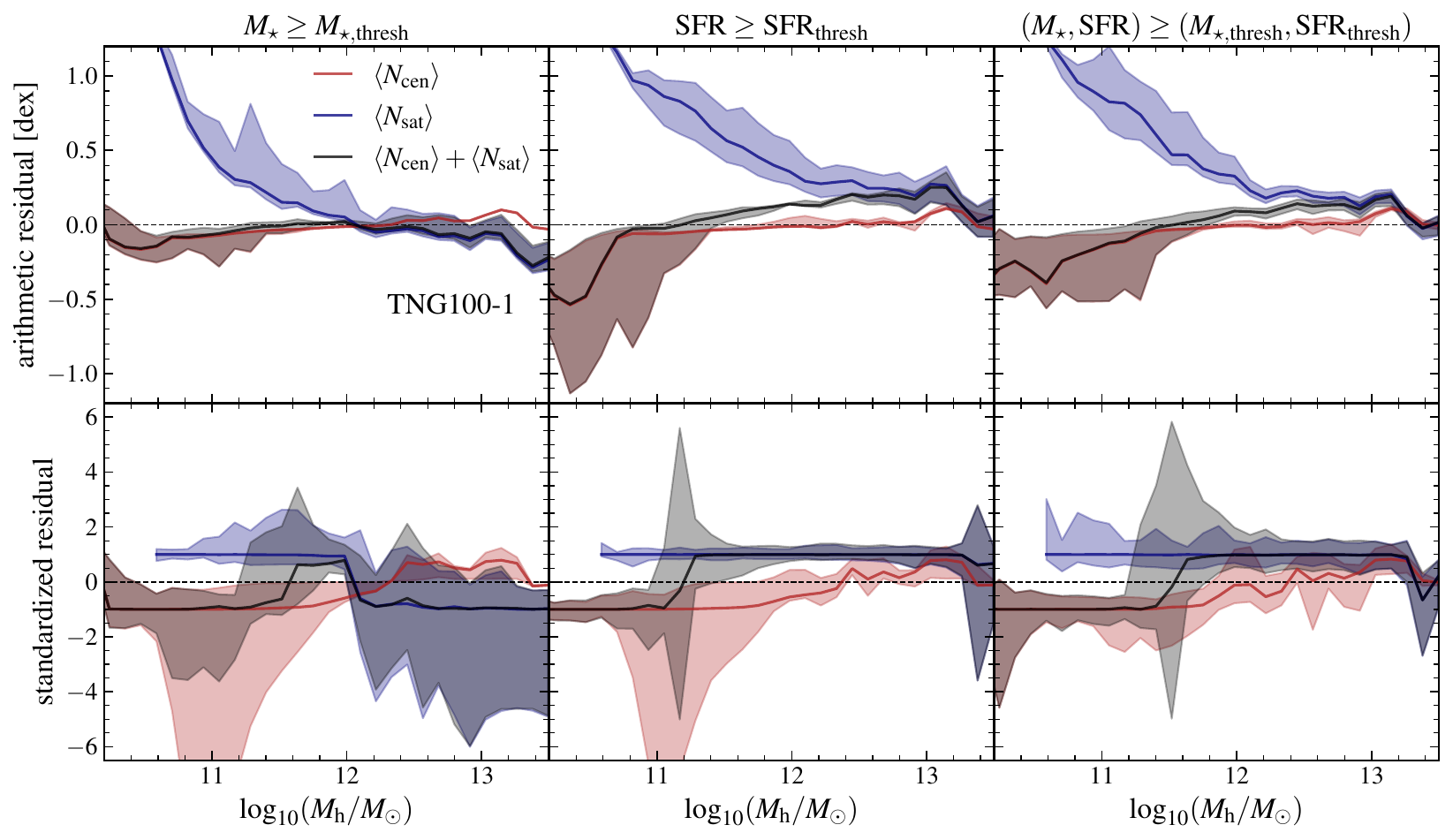}
\caption{Same as Figure \ref{fig:several_realization_nonparametric_HOD_posterior_frac_errs}, but now for the TNG100-1 mocks. The central HOD is consistently recovered within $\approx 0.2$ dex across the three different types of galaxy samples. The satellite HOD recovery performance is significantly worse than for the SC SAM mocks in Figure \ref{fig:several_realization_nonparametric_HOD_posterior_frac_errs}, especially at lower halo masses where the median arithmetic residual frequently exceeds 0.3 dex. To generate the standardized residual curves shown here, we rescale the 68\% confidence intervals of the recovered HOD posteriors by the nominal standardized residual in the HOD, producing median standardized residuals that are bounded between $\pm 1 \sigma$.}\label{fig:several_realization_nonparametric_HOD_posterior_frac_errs_TNG}
\end{figure*}

To evaluate the recovery performance of our methodology, we compute the arithmetic and standardized residuals in the recovered HODs for the test set of 300 TNG100-1 mock observations in an analogous manner to Section \ref{sec:sc_sam_validation} in Figure \ref{fig:several_realization_nonparametric_HOD_posterior_frac_errs_TNG}. The TNG100-1 mocks are generated using the same selection criteria tuples as the SC SAM validation set mocks in Section \ref{sec:sc_sam_validation}, and provide a non-circular assessment of the method's performance (given the emulator is trained on SC SAM mocks). To generate the standardized residual curves shown in Figure \ref{fig:several_realization_nonparametric_HOD_posterior_frac_errs_TNG}, we rescale the 68\% confidence intervals of the recovered HOD posteriors by the nominal standardized residual in the HOD, producing median standardized residuals that are bounded between $\pm 1\sigma$. Under the assumption that the level of offset between TNG100-1 and the SC SAM is similar to the offset between the real universe and the SC SAM, this makes our reported uncertainties a reasonable estimate for analysis of observational data. We demonstrate the non-parametric HOD recovery for a single TNG100-1 observation mock in Figure \ref{fig:nonparametric_HOD_posterior_TNG100-1}.

In Figure \ref{fig:several_realization_nonparametric_HOD_posterior_frac_errs_TNG}, there is a noticeable difference in overall HOD recovery performance for the TNG100-1 mocks relative to the SC SAM mocks in Figure \ref{fig:several_realization_nonparametric_HOD_posterior_frac_errs}. While the central HODs are consistently recovered within $\approx 0.2$ dex, our method consistently underestimates the satellite HOD by a significant margin ($\gtrsim 0.3$ dex) across nearly all types of galaxy samples. This behaviour is directly observable in Figure \ref{fig:nonparametric_HOD_posterior_TNG100-1}, where the recovered satellite HOD falls below the mock data points, ultimately producing a worse fit than for the SC SAM mock HOD in Figure \ref{fig:nonparametric_HOD_posterior}. We note however, that the halo mass range where the arithmetic residuals in the satellite HOD typically exceed $0.3$ dex corresponds to lower mass bins where the satellite contribution to the total HOD is insignificant, as evidenced by the reasonable arithmetic residuals for the total HOD in the same bins.

We understand the biases in HOD recovery to arise in part from the differing treatment of satellite galaxies in the SC SAM and TNG100-1. As is evident in Figure \ref{fig:mstar_sfr_cen_sat_TNG_SC_comp}, the $M_{\star}$-SFR distribution of satellites in the two simulations have fairly distinct shapes, that produce different galaxy populations when considering lower bound threshold-selected samples. Further, the radial distribution of satellites in the two simulations--which strongly influences the clustering in the one-halo regime of $w(\theta)$--differ as well (see Appendix \ref{sec:radial_distr_sats} for details). An additional source of disparity can be attributed to differences in the distribution of satellite multiplicities in the two simulations. The one-halo term of the $w(\theta)$ is sensitive to both $\langle N_{\mathrm{sat}} \rangle$ \textit{and} $\langle N_{\mathrm{sat}}^2 \rangle$ in Equation \eqref{eq:1h_full_term_final}, both of which are have different, and distinct, characteristics across the SC SAM and TNG100-1. We describe these differences in Appendix \ref{sec:sc_sam_tng_HOD_comparison}. The cumulative effect of these factors can be understood to modify the correspondence between $w(\theta)$ and the HOD that our method is reliant on, and the nature of which has been learned from the SC SAM. As a result, the recovery performance for satellite HODs in TNG100-1 is noticeably poorer than for satellite HODs in the SC SAM, and especially pronounced for galaxy samples selected using SFR (see Section \ref{sec:caveat_discussion} for further discussion).

In order to compare the HOD recovery performance of our non-parametric approach relative to the standard parametric routine, we similarly compute the arithmetic and standardized residuals for the parametric HOD recoveries for TNG100-1 mocks following the framework detailed in Appendix \ref{sec:appendix_parametric_HOD}, and adopting the \cite{Zheng_2005} and \cite{Geach2012} parameterizations. Given these parameterizations are often used to model the HODs of $M_{\star}$ and SFR-selected samples, the division of the 300 mocks into the three above-defined groups--only $M_{\star}$, only SFR, and joint ($M_{\star}$, SFR)--provides a natural basis for the comparison of HOD recovery performance across the non-parametric and parametric approaches. In Appendix \ref{sec:appendix_parametric_HOD}, we find that in general, our non-parametric HOD model recovers HODs with comparable or better precision and accuracy than parametric models across most halo mass bins.

\section{Discussion}\label{sec:caveat_discussion}

\subsection{Caveats of the SC SAM}

An in-depth discussion of the caveats pertaining to the SC SAM lightcone used in this paper is presented in \cite{10.1093/mnras/stac2139}. We draw attention to the SC SAM's treatment of satellite galaxies and the potential impact on the satellite HOD. In particular, assumptions embedded within the SC SAM may starve satellite galaxies of fresh gas, leading to premature quenching and reduced star formation. Given SFR is a primary selection criterion considered in this work, this process may reduce the satellite contribution to the HOD above a specified SFR threshold. Furthermore, the SC SAM implements a satellite position reassignment scheme in which the radial profile of satellites is adjusted to reproduce the dark matter density profile of the host dark matter halo (i.e.\ an NFW profile)--as is expected from $N$-body simulations. This has the potential to influence the one-halo term of the 2PCF, and subsequently influence the inferred satellite HOD. For these reasons, we will explore training our framework on alternative simulations in future work (see Section \ref{sec:future_work}). While training our routine on additional simulations should improve the overall generality of the framework, the ability to reconstruct HODs will still be limited by the range of HODs modelled in those simulations. 

\subsection{Caveats of the clustering model}

We also draw attention to the methodological limitations of the approach to modelling the one- and two-halo terms employed in this work. While our approximation is an efficient alternative to numerically estimating the 2PCF, it assumes a scale-independent galaxy bias. As a result, the approximation in Equation \eqref{eq:xi_efficient} tends to underestimate the two-halo contribution in the one-to-two-halo transition regime (i.e., $r \approx  [0.5, 3]$ cMpc or $\theta \approx [0.008, 0.04]$ deg for $z \sim 0.76$). To mitigate the impact of this on the HOD, we estimate a scale-dependent additive correction to the two-point correlation function of dark matter. This is calculated as
\begin{equation}
    w^{\text{corr}}_{\text{DM}}(\theta) = \left\langle \frac{w^{\text{treecorr}}(\theta) - w^{\text{1h}}(\theta)}{b_{\text{g}}^2} - w^{\text{lin}}_{\text{DM}}(\theta) \right\rangle
\end{equation}
where $w^{\text{treecorr}}(\theta)$ is the angular galaxy 2PCF estimated using \texttt{treecorr}, $w^{\text{1h}}(\theta)$ is the one-halo term in angular space, $b_{\text{g}}$ is the scale-independent galaxy bias, $w^{\text{lin}}_{\mathrm{DM}}(\theta)$ is the angular 2PCF of dark matter, and $\langle \dots \rangle$ represents a statistical average across the set of 300 SC SAM mock observations analyzed in Section \ref{sec:sc_sam_validation}. This serves as a semi-empirical correction to better match mock galaxy 2PCFs with those of the SC SAM model. The form of the modified angular 2PCF of dark matter used in this analysis is then
\begin{equation}
    w^{\text{lin, mod}}_{\text{DM}}(\theta) = w^{\text{lin}}_{\text{DM}}(\theta) + w^{\text{corr}}_{\text{DM}}(\theta).
\end{equation}

We also note that the linear halo bias $b_{\mathrm{h}}(M_{\mathrm{h}})$ in Equation \eqref{eq:xi_efficient} assumes a purely halo mass-dependent relation. While this is a common assumption in the literature (e.g., \citealt{Tinker_2010}), this formulation implicitly assumes the absence of \textit{galaxy assembly bias}--the dependence of the HOD on halo properties other than mass on which halo clustering \textit{also} depends \citep{2014MNRAS.443.3044Z}. Many studies have found that assembly bias is an important ingredient in determining the strength of galaxy clustering (e.g., \citealt{2007MNRAS.374.1303C}). Notably, \citet{2014MNRAS.443.3044Z} found that ignoring assembly bias in HOD inference introduces significant systematic errors. Incorporating assembly bias or avoiding the explicit use of the halo bias model altogether is left to future work. Because our one-halo term is empirically derived from simulations that implicitly account for assembly bias, our vulnerability to its exclusion is primarily on large scales.

\subsection{Future Extensions:  Additional Simulations, Survey Data, and Physical Galaxy Properties}\label{sec:future_work}

Future work will focus on expanding the applicability of our non-parametric HOD framework to alternative semi-analytic and hydrodynamical simulations (e.g.\ Horizon Run 5; \citealt{Lee_2021}, OpenUniverse2024; \citealt{openuniverse2025openuniverse2024sharedsimulatedview}). Training and testing our method on different simulations will improve its overall robustness and ensure that we are not biased to any single simulation suite. The OpenUniverse2024 simulations are particularly relevant given they are designed to simulate the overlap of Vera C. Rubin Observatory's Legacy Survey of Space and Time Wide-Fast-Deep survey (\citealt{Ivezic_2019}) and the \textit{Roman} High-Latitude Wide-Area survey \citep{Wang_2022}. Training our approach on these simulations will prepare our method to analyze the eventual data from these two surveys. 

We also intend to validate our non-parametric HOD method against real survey data by applying our methodology to a set of existing galaxy catalogues (e.g.,\ eBOSS; \citealt{Alam_2021}, HETDEX; \citealt{Gebhardt_2021}, DESI; \citealt{desicollaboration2025desidr2resultsi}). For spectroscopic galaxy surveys such as HETDEX and DESI, this will require the incorporation of redshift space distortions into our clustering analysis, enabling the analysis of the redshift-space 2PCF $\xi(s, \mu)$ \citep{1992ApJ...385L...5H, Hamilton_1998}.

In this work, we have used selection thresholds in stellar mass and star formation rate (as  a tuple of $M_\star$, SFR) as nuisance parameters to describe simulated galaxy samples.  Observed galaxy samples are most frequently selected to exceed a given value of flux density (i.e., apparent magnitude) in a chosen filter or wavelength.  This does not create a direct correspondence with a physical property such as stellar mass, even when the chosen filter falls in the rest-frame near-infrared where dust attenuation is minimal, due to observational incompleteness that is strongest near the selection limit. Inferring the distribution of physical galaxy properties from the observational flux density limit is difficult due to the challenges of forward modelling the galaxies' dust and metallicity and their effects on observed photometry.  This suggests a future extension of our method to simultaneously infer both the HOD and the distribution of underlying physical properties that correspond to an observed galaxy sample.  The mock galaxy catalog offers a long list of physical properties; if they match the real universe well, our machinery can be used for this simultaneous inference. 

\section{Conclusions}\label{sec:conclusion}

The statistical relationship between the galaxies we observe and their host dark matter haloes is a product of both the underlying cosmology and the complex physics of galaxy formation and evolution. HODs present a computationally inexpensive framework to empirically quantify the galaxy-halo connection from measurements of galaxy clustering. However, standard parametric models lack the necessary flexibility to describe the complete space of physically plausible galaxy samples that are anticipated from current and next-generation galaxy surveys.

In this work, we have introduced a non-parametric approach to HOD modelling designed to resolve the issues associated with parametric modelling routines. In this setting, we did not assume a functional form for the HOD. We trained a Gaussian process-based algorithm to learn the mappings among the physical galaxy property space, the galaxy 2PCF space, \textit{and} the HOD space using a single suite of SC SAM lightcones. Incorporating this emulator into a Bayesian inference framework, we derived simultaneous constraints on physical galaxy property limits and HODs when provided with a mock measurement of the galaxy 2PCF. In this initial implementation, our non-parametric HOD framework reliably recovers the central, satellite, and total HODs within 0.1 dex for the SC SAM validation set and consistently within 0.2 dex for the TNG100-1 test set, and with accurate uncertainties that have been tuned to estimate those expected in the analysis of observational data. 

Using a parametric HOD recovery with the \cite{Zheng_2005} and \cite{Geach2012} parameterizations as benchmark models, we find that our non-parametric HOD model recovers HODs with comparable or better precision and accuracy than parametric inference across most halo mass bins. Our non-parametric HOD approach also benefits from a substantial reduction in computational complexity provided by the GP-based emulator, reducing the average time for a likelihood evaluation.

The development of a non-parametric method for HOD inference  is a crucial step toward fully leveraging the statistical power of current and next-generation galaxy surveys. As observational data become increasingly precise and diverse—probing a wider range of galaxy populations, environments, and redshifts—the limitations of rigid parametric models will become more pronounced. These models, by construction, embed assumptions about the form of the galaxy-halo connection that may not hold across different galaxy selections or cosmological regimes. A flexible approach, such as that introduced in this paper, therefore becomes essential for capturing the diversity and complexity of real galaxy populations.

Robust modelling of the galaxy-halo connection is particularly important when using large-scale structure data for cosmological or astrophysical inference. Inaccurate or oversimplified HOD assumptions can propagate into biased constraints on the quantities of interest. By avoiding strong prior assumptions on the shape of the HOD, our non-parametric HOD framework provides a more reliable and interpretable foundation for such analyses. 

\begin{acknowledgments}
The authors would like to thank David Weinberg and Joshua Kable for valuable comments on the manuscript.  
JK and EG acknowledge support from NSF grant AST-2206222 and from the U.S. Department of Energy, Office of Science, Office of High Energy Physics Cosmic Frontier Research program under Award Number DE-SC0010008. Support for KI was provided by NASA through the NASA Hubble Fellowship grant HST-HF2-51508 awarded by the Space Telescope Science Institute, which is operated by the Association of Universities for Research in Astronomy, Inc., for NASA, under contract NAS5-26555. AY is supported by a Giacconi Fellowship from the Space Telescope Science Institute, which is operated by the Association of Universities for Research in Astronomy, Incorporated, under NASA contract HST NAS5-26555 and JWST NAS5-03127. JK and EG acknowledge the Office of Advanced Research Computing (OARC) at Rutgers, The State University of New Jersey for providing access to the Amarel cluster and associated research computing resources that have contributed to the results reported here.

\end{acknowledgments}

%

\vspace{5mm}


\software{Python, astropy, pyCCL, treecorr, GPy, halomod }



\appendix

\section{Parametric HOD modelling}\label{sec:appendix_parametric_HOD}

To demonstrate the improvement in HOD modelling that our non-parametric approach may provide relative to the standard parametric routine, we apply parametric HOD modelling using the \cite{Zheng_2005} and \cite{Geach2012} parameterizations to the same set of 300 TNG100-1 mock observations that we test our non-parametric approach on in Section \ref{sec:tng_results}. To perform this investigation, we make use of the Python package \texttt{halomod}\footnote{\url{https://github.com/halomod/halomod},v2.2} \citep{murray2013hmfcalconlinetoolcalculating, Murray_2021}. \texttt{halomod} enables efficient HOD modelling that can be easily interfaced with a MCMC framework to constrain the parameters of a given HOD parameterization from a measurement of the 2PCF. In this work, we consider two HOD parameterizations; (1) the five-parameter model introduced in \cite{Zheng_2005} designed to describe stellar mass-selected galaxy samples, and (2) the eight-parameter model introduced in \cite{Geach2012} designed to describe SFR-selected galaxy samples. The five-parameter \citet{Zheng_2005} model has central and satellite HODs of the form
\begin{align}
    \langle N_{\text{cen}}(M_{\text{h}}) \rangle &= \frac{1}{2} \left[1+ \text{erf} \left(\frac{\log_{10}(M_{\text{h}}) - \log_{10}(M_{\text{min}})}{\sigma_{\log M}} \right) \right] \label{eq:zheng_cen_HOD} \\
    \langle N_{\text{sat}}(M_{\text{h}}) \rangle &= \langle N_{\text{cen}}(M_{\text{h}}) \rangle \Theta(M_{\text{h}} - M_{\text{cut}}) \left(\frac{M_{\text{h}} - M_{\text{cut}}}{M_1}\right)^{\alpha} \label{eq:zheng_sat_HOD}
\end{align}
where $\text{erf}$ is the Gauss error function, $M_{\text{min}}$ is the minimum halo mass that hosts galaxies, and $\sigma_{\text{log}M}$ is the characteristic transition width, $\Theta$ is the Heaviside function, $M_{\text{cut}}$ is the minimum halo mass that hosts satellites, $M_{1}$ is a normalization factor, and $\alpha$ is the power law index. In Equation \eqref{eq:zheng_sat_HOD}, we have chosen the common version of the \cite{Zheng_2005} parameterization that conditions the satellite HOD on the central HOD. This ensures that when $\langle N_{\mathrm{cen}} (M_{\mathrm{h}}) \rangle = 0$, $\langle N_{\mathrm{sat}} (M_{\mathrm{h}}) \rangle=0$ as well. The eight-parameter \cite{Geach2012} has a central HOD
\begin{equation}
    \begin{aligned}
        \langle N_{\text{cen}}(M_{\text{h}}) \rangle  &= F_c^B (1-F_c^A) \, \text{exp} \left[-\frac{(\log_{10}(M_{\text{h}}) - \log_{10}(M_c))^2}{2 \beta_{\log M}^2}\right] \\
        &\quad + F_c^A \left[1+\text{erf}\left(\frac{\log_{10}(M_{\text{h}}) - \text{log}(M_c)}{\beta_{\log_{10} M}} \right) \right]
    \end{aligned}
    \label{eq:geach_cen_HOD}
\end{equation}
where $F_c^{A, B}$ are normalization factors ranging from 0 to 1, $M_c$ is the mean of the Gaussian distribution and low-mass cut-off for the step-function, and $\beta_{\text{log}M}$ is the width of the Gaussian and transition width of the step-function. The \cite{Geach2012} satellite HOD has the form 
\begin{equation}
    \begin{aligned}
        \langle N_{\text{sat}}(M_{\text{h}}) \rangle  &= F_{\text{s}} \left( \frac{M_{\text{h}}}{M_{\text{min}}} \right)^{\gamma} \times \\ &  \left[1+\text{erf}\left(\frac{\log_{10}(M_{\text{h}}) - \log_{10}(M_{\text{min}})}{\delta_{\log M}} \right) \right].
    \label{eq:geach_sat_HOD}
    \end{aligned}
\end{equation}
where $M_{\text{min}}$ is the minimum mass of haloes that host satellites, $F_{\text{s}}$ is the mean number of galaxies at $M_{\text{min}}$, $\delta_{\text{log}M}$ is satellite transition scale,  and $\gamma$ is the power law index. 

To constrain the HOD from a measurement of the 2PCF using either of the above parameterizations, we perform HOD parameter space exploration using the MCMC package \texttt{emcee} \citep{Foreman_Mackey_2013}. This involves constructing a forward model of the 2PCF as a function of the HOD parameters. To do so, we make use of the \texttt{halo\_model} module in $\texttt{halomod}$. We use the \texttt{halo\_model} module to compute the one-halo term of the galaxy 2PCF using Equation \eqref{eq:1h_full_term_final}, and the two-halo term using the linear large-scale galaxy bias approximation in Equation \eqref{eq:2halo_term}. This formulation of the one- and two-halo terms is intended to improve the agreement between our approach to estimating $w(\theta)$ detailed in Section \ref{sec:sc_sam_sims}. We specify the normalized radial number density profile of satellites $u_{\mathrm{s}}(r| M_{\mathrm{h}})$ to be a NFW profile \citep{1996ApJ...462..563N}, as is expected for satellites in TNG100-1 \citep{nelson2021illustristngsimulationspublicdata}. We implement the same MCMC configuration as detailed in Section \ref{sec:MCMC}, albeit with a modified posterior of the form
\begin{equation}
    p(\boldsymbol{\theta_{\text{HOD}}} | w^{\text{obs}}(\theta)) \propto p(w^{\text{obs}}(\theta) | \boldsymbol{\theta_{\text{HOD}}}) p (\boldsymbol{\theta_{\text{HOD}}})
\end{equation}
where $w^{\text{obs}}$ is the mock observation of the 2PCF being analyzed and $\boldsymbol{\theta}_{\text{HOD}}$ are the set of HOD parameters corresponding to either of the \cite{Zheng_2005} or \cite{Geach2012} parameterizations. Adopting a Gaussian likelihood for the 2PCF, a Gaussian number density penalty (analogous to that detailed in Section \ref{sec:MCMC}), and a uniform top-hat prior over the HOD parameters, the logarithm of the posterior the MCMC aims to maximize is
\begin{equation}
\begin{aligned}
    \ln p(w^{\text{obs}}(\theta) | \boldsymbol{\theta_{\text{HOD}}})  & \propto \\ -\frac{1}{2} \left(w^{\text{mod}}(\theta) - w^{\text{obs}}(\theta) \right)^{\text{T}} \mathbf{C^{\text{obs}}}^{-1} \left(w^{\text{mod}}(\theta) - w^{\text{obs}}(\theta) \right) \\ - \frac{\left(n_{\text{g, mod}} - n_{\text{g, obs}}\right)^2}{2\sigma^2_{n_{\text{g, obs}}}}  \\ - \sum_{i=1}^{N_{\theta}} \ln \Theta(\boldsymbol{\theta}_i - \boldsymbol{\theta}_{i, \text{min}})+ \ln \Theta(\boldsymbol{\theta}_{i, \text{max}}-\boldsymbol{\theta}_i)
\end{aligned}\label{eq:parametric_posterior}
\end{equation}
where $\mathbf{C^{\text{obs}}}$ is the covariance matrix of the mock observation (computed using jackknife resampling), $w^{\text{mod}}$ is the 2PCF computed at $\theta_{\text{HOD}}$, $n_{\text{g, mod}}$ is the galaxy number density corresponding to $w^{\text{mod}}$, $n_{\text{g, obs}}$ is the galaxy number density corresponding to $w^{\text{obs}}$, and $\sigma^2_{n_{\text{g, obs}}}$ is the variance in $n_{\text{g, obs}}$. The final term enforces the uniform top-hat prior on the HOD parameters, where $N_{\theta}$ is the number of parameters, $\boldsymbol{\theta}_{i, \text{min}}$ and $\boldsymbol{\theta}_{i, \text{max}}$ denote the minimum and maximum values of the top-hat prior for the $i$th component of $\boldsymbol{\theta_{\text{HOD}}}$, respectively. The values of the parameter bounds used for each of the \cite{Zheng_2005} and \cite{Geach2012} parameterizations are listed in Tables \ref{tab:zheng_MCMC_bounds_parametric} and \ref{tab:geach_MCMC_bounds_parametric}, respectively.

\begin{table}
    \centering
    \begin{tabular}{ll}
        \hline
        \textbf{Parameter} & \textbf{Prior} \\
        \hline
        $\log_{10}(M_\mathrm{{min}})$ & $\mathcal{U}(9.0, 13.0)$ \\
        $\log_{10}(M_\mathrm{1})$ & $\mathcal{U}(9.0, 13.5)$ \\
        $\alpha$ & $\mathcal{U}(0, 2.0)$ \\
        $\sigma_{\log M}$ & $\mathcal{U}(0, 0.5)$ \\
        $\log_{10}(M_\mathrm{cut})$ & $\mathcal{U}(9.0, 13.5)$ \\
        \hline
    \end{tabular}
    \caption{Uniform top-hat priors for the \cite{Zheng_2005} parameters.}
    \label{tab:zheng_MCMC_bounds_parametric}
\end{table}

\begin{table}[h]
    \centering
    \begin{tabular}{ll}
        \hline
        \textbf{Parameter} & \textbf{Prior} \\
        \hline
        $\log_{10}(M_c)$ & $\mathcal{U}(9.0, 13.0)$ \\
        $\log_{10}(M_\mathrm{min})$ & $\mathcal{U}(9.0, 13.5)$ \\
        $\gamma$ & $\mathcal{U}(0, 2.0)$ \\
        $\beta_{\log M}$ & $\mathcal{U}(0, 0.5)$ \\
        $F^A_c$ & $\mathcal{U}(0, 1.0)$ \\
        $F^B_c$ & $\mathcal{U}(0, 1.0)$ \\
        $F_s$ & $\mathcal{U}(0, 1.0)$ \\
        $\delta_{\log M}$ & $\mathcal{U}(0, 0.5)$ \\
        \hline
    \end{tabular}
    \caption{Uniform top-hat priors for the \cite{Geach2012} parameters.}
    \label{tab:geach_MCMC_bounds_parametric}
\end{table}

\begin{figure*}[htbp]
\centering
\includegraphics[width=0.80\textwidth]{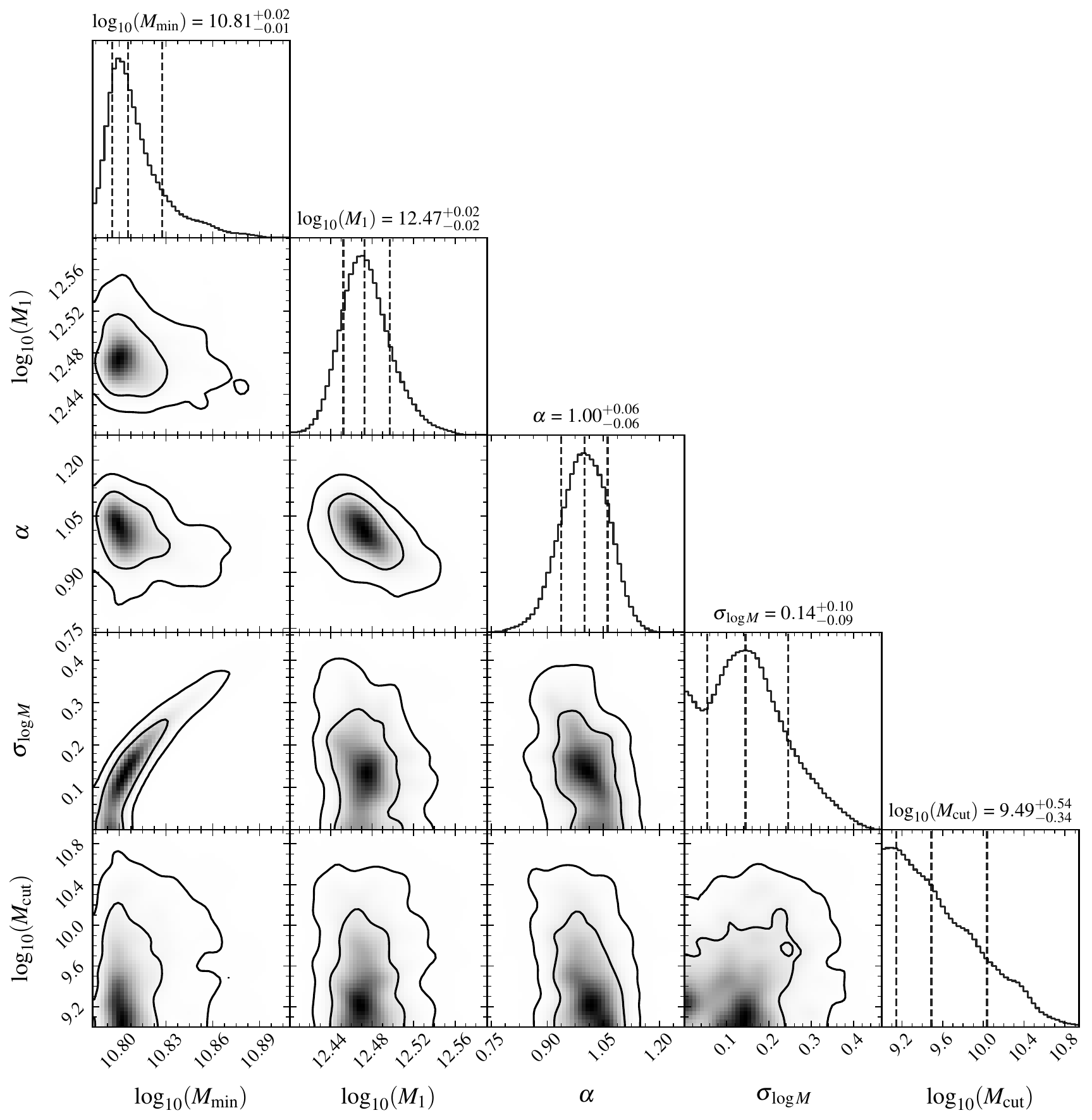}
\caption{The MCMC corner plot for the \cite{Zheng_2005} parameterization using a TNG100-1 mock 2PCF observation with $(\log_{10}(M_{\star}/M_{\odot}), \log_{10}(\mathrm{SFR}/M_{\odot} \mathrm{yr}^{-1})) 
\geq (8.45, -2.45)$ (i.e., the same input 2PCF as used in Figure \ref{fig:nonparametric_corner_plot}). The black lines in the one-dimensional posteriors indicate the 16th (left dashed), 50th (middle dashed), 
and 84th (right dashed) percentiles. The inner and outer contours in the two-dimensional posterior contain 68\% and 95\% of the 
samples, respectively.  Because the galaxy 2PCFs we are fitting were generated from a non-parametric HOD, no ``true" set of HOD parameters is available to plot.}\label{fig:zheng_corner_plot_example}
\end{figure*}

\begin{figure*}[htbp]
\includegraphics[width=\textwidth]{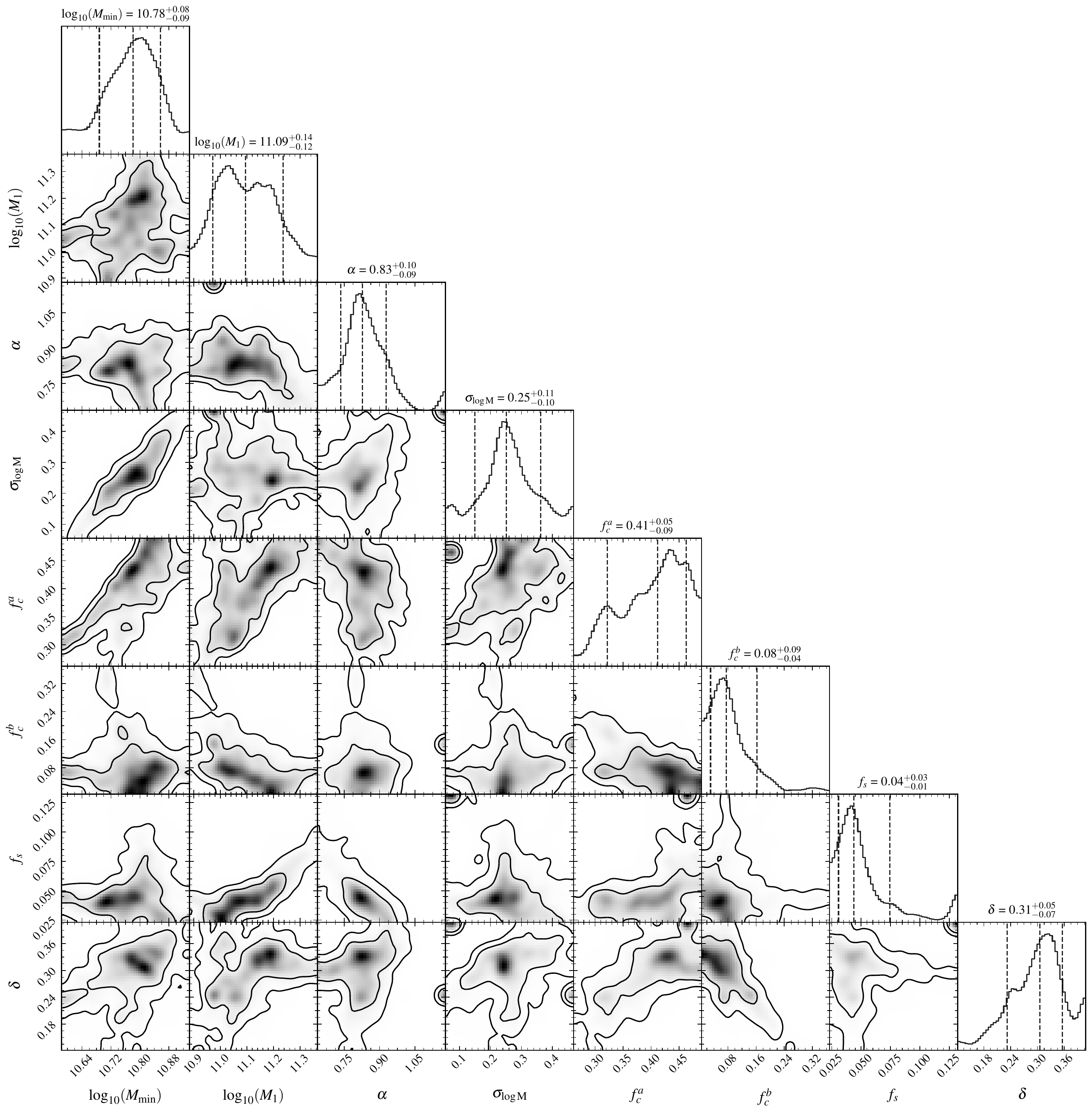}
\caption{The same as Figure \ref{fig:zheng_corner_plot_example}, but now for the \cite{Geach2012} parameterization, with the same input TNG100-1 mock 2PCF observation.} \label{fig:geach_corner_plot_example}
\end{figure*}

\begin{figure}[htbp]
\includegraphics[width=\columnwidth]{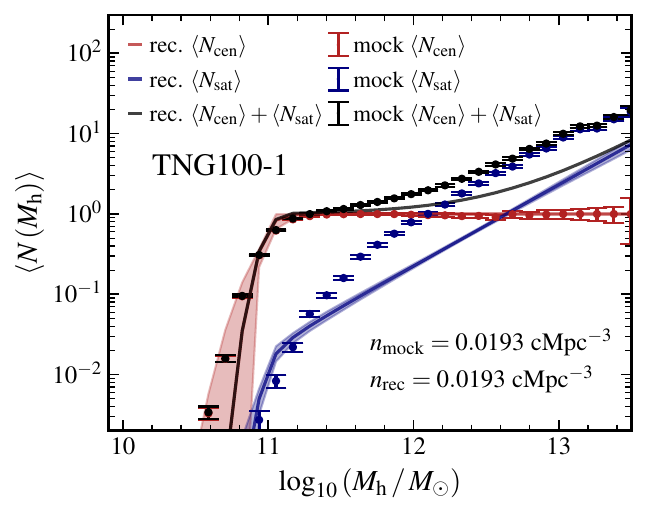}
\caption{The median recovered central (solid red), satellite (solid blue), and total (solid black) HODs corresponding to the corner plot in Figure \ref{fig:zheng_corner_plot_example} for the \cite{Zheng_2005} parameterization. The 68\% confidence intervals for each curve are indicated as shaded regions. The ``true" HODs of the mock observation--the same as in Figure \ref{fig:nonparametric_HOD_posterior_TNG100-1}--are shown as data points. The \cite{Zheng_2005} parameterization closely reproduces the central HOD, yet significantly underestimates the satellite HOD of the TNG100-1 mock.}
\label{fig:zheng_HOD_posterior}
\end{figure}

\begin{figure}[htbp]
\includegraphics[width=\columnwidth]{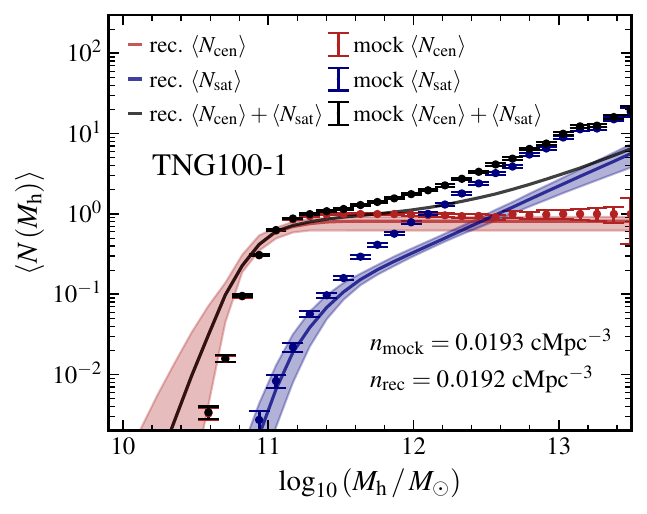}
\caption{The same as Figure \ref{fig:zheng_HOD_posterior}, but now for the \cite{Geach2012} parameterization. The \cite{Geach2012} parameterization consistently underestimates both the central HOD and the satellite HOD for $\log_{10}(M_{\mathrm{h}}/M_{\odot}) \gtrsim 11$.}
\label{fig:geach_HOD_posterior}
\end{figure}

\begin{figure*}[htbp]
\includegraphics[width=\textwidth]{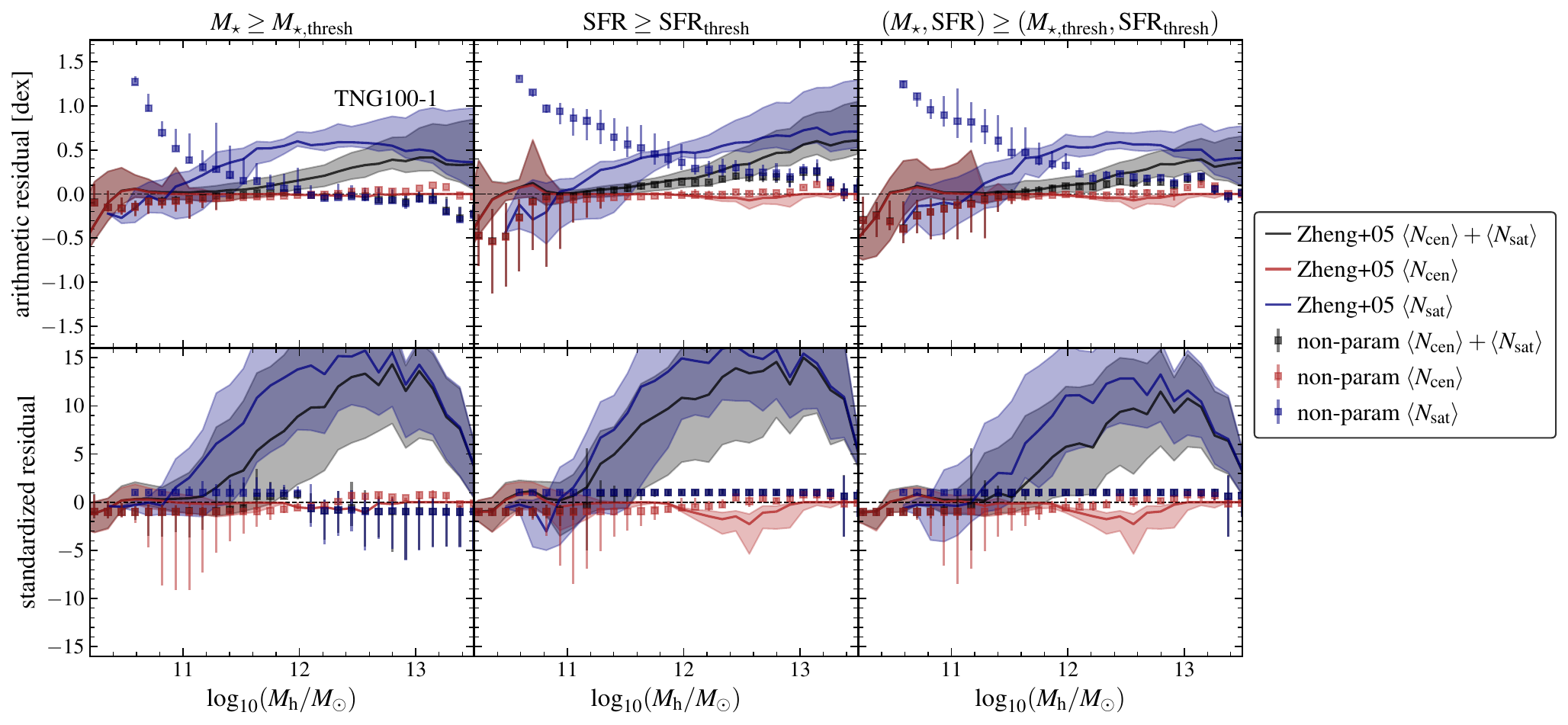}
\caption{Same as Figure \ref{fig:several_realization_nonparametric_HOD_posterior_frac_errs_TNG}, although now using the parametric HOD modelling routine with the \cite{Zheng_2005} parameterization. The arithmetic and standardized residuals corresponding to the non-parametric HOD recoveries over the TNG100-1 mocks are plotted as data points for direct comparison with the parametric recovery curves. The errorbars on the non-parametric data points represent the 68\% confidence intervals (previously represented as shaded regions in Figure \ref{fig:several_realization_nonparametric_HOD_posterior_frac_errs_TNG}).} \label{fig:several_realization_zheng_parametric_HOD_posterior_frac_errs}
\end{figure*}

\begin{figure*}[htbp]
\includegraphics[width=\textwidth]{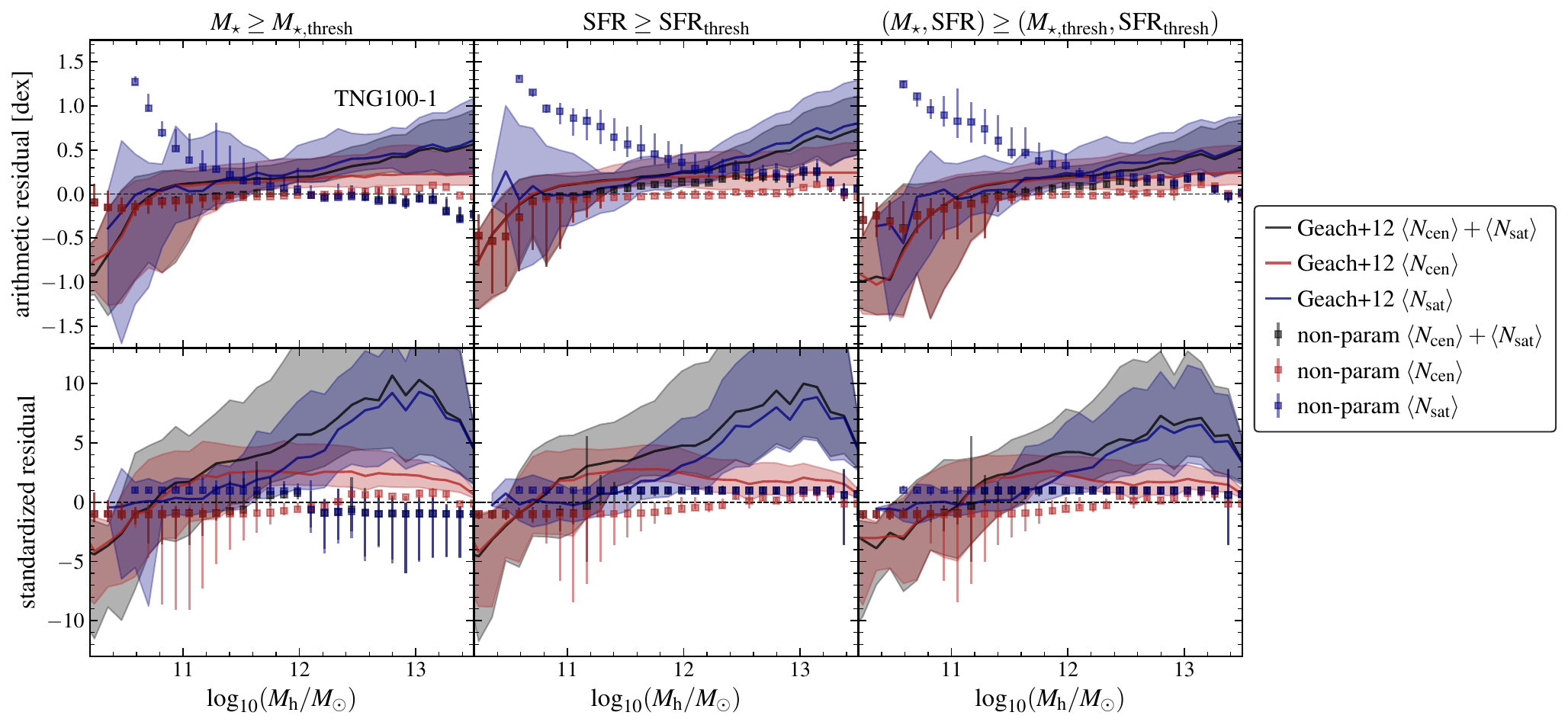}
\caption{The same as Figure \ref{fig:several_realization_zheng_parametric_HOD_posterior_frac_errs}, although now using the parametric HOD modelling routine with the \cite{Geach2012} parameterization.} \label{fig:several_realization_geach_parametric_HOD_posterior_frac_errs}
\end{figure*}

As a baseline comparison for our non-parametric HOD model, we run MCMC recoveries using both the \cite{Zheng_2005} and \cite{Geach2012} parameterizations for the 300 TNG100-1 mock observations discussed in Section \ref{sec:tng_results}. Equipped with the parameter chains corresponding to each mock observation, we randomly draw 1000 samples and compute the median central, satellite and total HOD and their respective 68\% confidence intervals. We then characterize the parametric HOD recovery performance by computing the median and 68\% confidence intervals for both the arithmetic and standardized residuals as done for the non-parametric HOD recoveries conducted in Section \ref{sec:tng_results}. In Figures \ref{fig:zheng_corner_plot_example}, \ref{fig:geach_corner_plot_example}, \ref{fig:zheng_HOD_posterior}, and \ref{fig:geach_HOD_posterior} we show the MCMC corner plots and corresponding HOD posteriors for both the \cite{Zheng_2005} and \cite{Geach2012} parameterizations when provided with the same TNG100-1 mock observation used in Figure \ref{fig:nonparametric_HOD_posterior_TNG100-1}. In Figures \ref{fig:zheng_HOD_posterior} and \ref{fig:geach_HOD_posterior}, the recovered ``best-fit" parametric HODs generally fail to capture the unique shape of the central and satellite HODs. Figures \ref{fig:several_realization_zheng_parametric_HOD_posterior_frac_errs} and \ref{fig:several_realization_geach_parametric_HOD_posterior_frac_errs} demonstrate the statistical behaviour of the HOD recovery performance for the \cite{Zheng_2005} and \cite{Geach2012} parameterizations, respectively. In Figure \ref{fig:several_realization_zheng_parametric_HOD_posterior_frac_errs}, the \citet{Zheng_2005} parametric recoveries of $M_{\star}$-selected samples exhibit comparable arithmetic residuals to our non-parametric recoveries in the central HOD, and larger arithmetic residuals in the satellite HOD (excluding $\log_{10}(M_{\mathrm{h}}/M_{\odot}) \lesssim 11$). For SFR- and joint ($M_{\star}$, SFR)-selected samples in \ref{fig:several_realization_zheng_parametric_HOD_posterior_frac_errs}, the \citet{Zheng_2005} parametric recoveries have comparable or larger arithmetic residuals to our non-parametric recoveries in the central HOD and in the satellite HOD for $\log_{10}(M_{\mathrm{h}} / M_{\odot}) \gtrsim 11.5$. In Figure \ref{fig:several_realization_zheng_parametric_HOD_posterior_frac_errs}, the standardized residuals in the central HOD for the \citet{Zheng_2005} parametric recoveries are reasonably consistent with our non-parametric recoveries, however, the standardized residuals for the satellite HOD are heavily biased for $\log_{10}(M_{\mathrm{h}}/M_{\odot}) \gtrsim 11$ at $\gtrsim +3\sigma$, indicating that the predicted uncertainties for the \citet{Zheng_2005} satellite HODs are underestimated. In Figure \ref{fig:several_realization_geach_parametric_HOD_posterior_frac_errs}, the \citet{Geach2012} parametric recoveries of $M_{\star}$-selected samples exhibit comparable or larger arithmetic residuals to our non-parametric recoveries in the central HOD and the satellite HOD for $\log_{10}(M_{\mathrm{h}} / M_{\odot}) \gtrsim 11$. For SFR- and joint ($M_{\star}$, SFR)-selected samples in \ref{fig:several_realization_geach_parametric_HOD_posterior_frac_errs}, the \citet{Geach2012} parametric recoveries have comparable or larger arithmetic residuals to our non-parametric recoveries in the central HOD and in the satellite HOD for $\log_{10}(M_{\mathrm{h}} / M_{\odot}) \gtrsim 11.5$. In Figure \ref{fig:several_realization_geach_parametric_HOD_posterior_frac_errs}, the standardized residuals in the central and satellite HODs for the \citet{Geach2012} parametric recoveries are biased towards overestimating the HOD ($\approx +2\sigma$ for $\langle N_{\mathrm{cen}} \rangle$ and $\gtrsim +1\sigma$ for $\langle N_{\mathrm{sat}} \rangle$).

\section{Radial distribution of satellites}\label{sec:radial_distr_sats}

\begin{figure*}[htbp]
\centering
\includegraphics[width=0.9\textwidth]{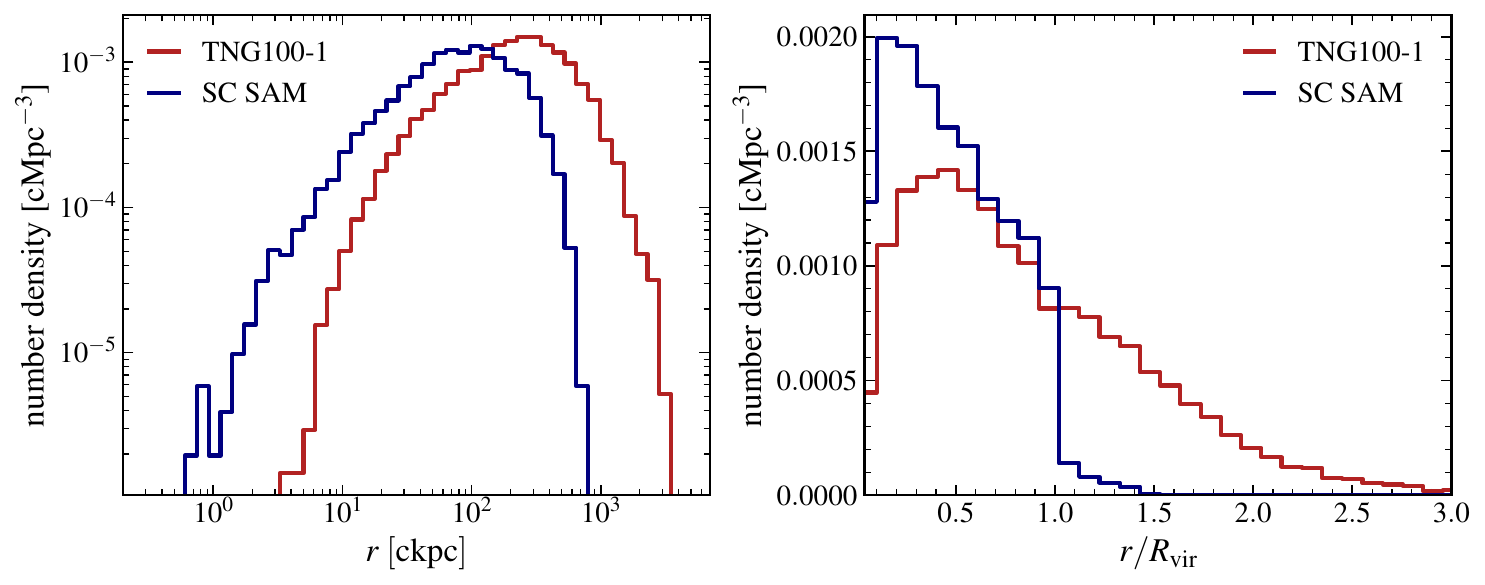}
\caption{Left panel: Distribution of central-satellite separations for galaxies with ($\log_{10}(M_{\star}/M_{\odot})$,  $\log_{10}(\mathrm{SFR} / M_{\odot} \mathrm{yr}^{-1})) \geq (7, -3)$ in TNG100-1 (red) and the SC SAM (blue). The TNG100-1 distribution is shifted towards larger separations, reaching a maximum at $r\approx 4 \times 10^2$ ckpc, whilst the SC SAM distribution reaches a maximum at $r\approx 10^2$ ckpc. Right panel: The same as the left panel, although now normalizing the central-satellite separations by the host halo virial radius. The SC SAM exhibits a distinctive drop-off around $r/R_{\mathrm{vir}} \sim 1.0$, while TNG declines steadily over the range $r/R_{\mathrm{vir}} \sim [0.5, 3.0]$.}\label{fig:radial_dist_of_satellites}
\end{figure*}

The distribution of central-satellite separations in the SC SAM and TNG100-1 are shown in the left panel of Figure \ref{fig:radial_dist_of_satellites}. Notably, the SC SAM has a maxima that is $\approx 4$x smaller than TNG100-1. Consequently, the distribution of host halo virial radius normalized central-satellite separations shown in the right panel of Figure \ref{fig:radial_dist_of_satellites} indicates that the radial profile of satellites falls off much more steeply in the SC SAM than in TNG100-1, which exhibits a steady decline out to $r/R_{\mathrm{vir}} \approx 3.0$. We believe that the discrepancies between the radial distribution of satellites in haloes in the SC SAM and TNG100-1, coupled with the enhanced quenching of satellites in the SC SAM (as described in Section \ref{sec:caveat_discussion}), are responsible for the marked difference in non-parametric HOD recovery performance between the SC SAM and TNG100-1 mocks, given that they modify the nominal $w(\theta)$-HOD correspondence our method has identified.

\section{SC SAM and TNG100-1 HOD Comparison}\label{sec:sc_sam_tng_HOD_comparison}

\begin{figure*}
\centering
\includegraphics[width=0.75\textwidth]{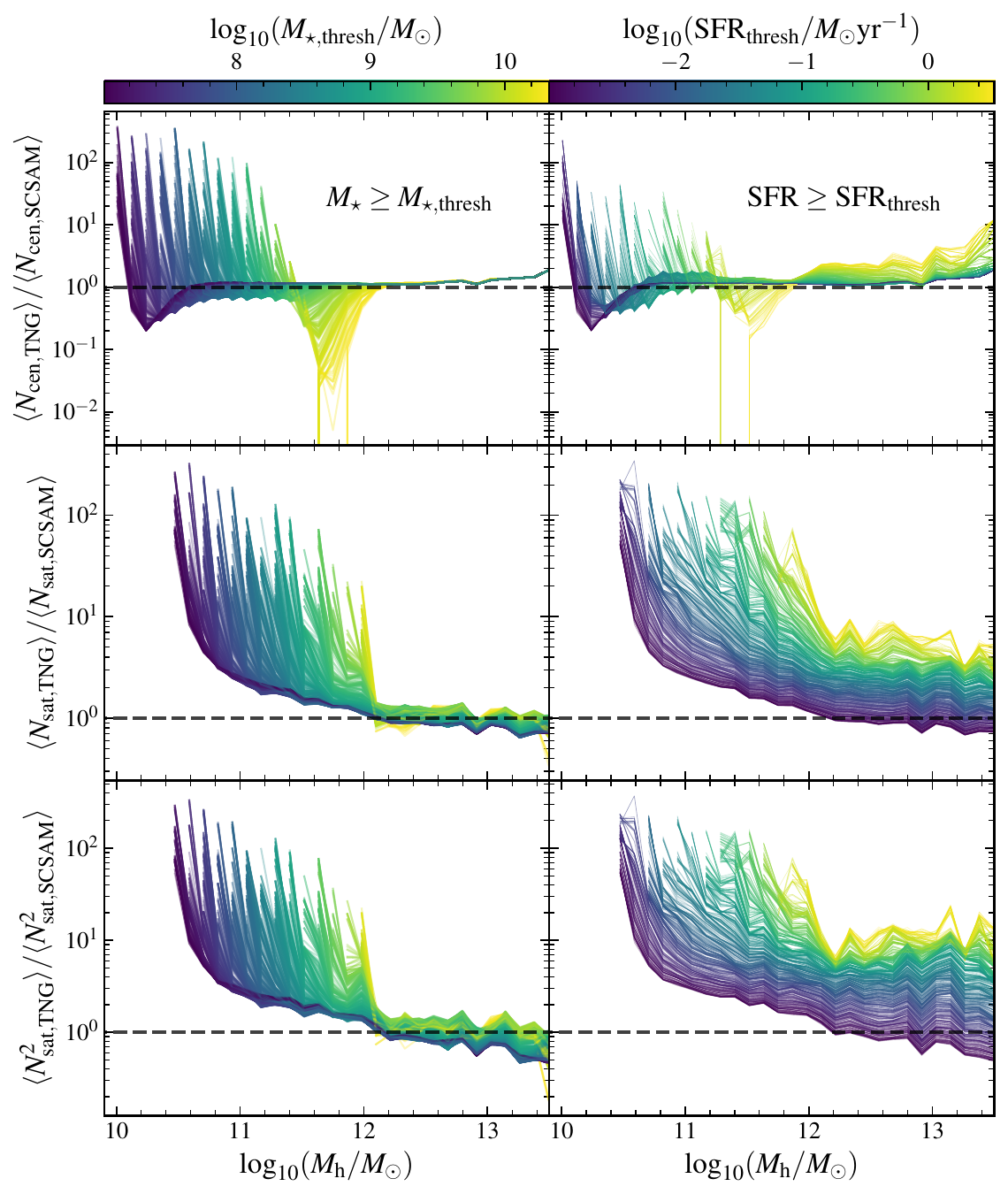}
\caption{Left column: The ratios between TNG100-1 and SC SAM central HODs (upper row), satellite HODs (middle row), and $\langle N_{\mathrm{sat}}^2 \rangle$ (bottom row), for variable lower bound thresholds of $M_{\star}$. Right column: Same as the left column, but now for variable lower bound thresholds of SFR. The range of amplitudes covered by the curves is indicative of the substantial differences in HODs between the two simulations.} \label{fig:sc_sam_tng_HOD_comp}
\end{figure*}

Figure \ref{fig:sc_sam_tng_HOD_comp} displays the ratio between the TNG100-1 and SC SAM $\langle N_{\mathrm{cen}} \rangle$, $\langle N_{\mathrm{sat}} \rangle$, and $\langle N_{\mathrm{sat}}^2 \rangle$ for the range of $M_{\star}$ and SFR lower bound thresholds considered in this work. Notably, there are significant differences in the HODs of the two simulations for the same lower bound thresholds. The discrepancy is especially significant at low-to-intermediate halo masses (e.g., $\log_{10}(M_{\star}/M_{\odot}) \lesssim 12$), and generally decreases as halo mass increases. In particular, the differences in magnitude of the $\langle N_{\mathrm{cen}} \rangle$, $\langle N_{\mathrm{sat}} \rangle$, and $\langle N^2_{\mathrm{sat}}\rangle$ across the two simulations increases as the lower bound threshold on SFR increases, indicating that the corresponding galaxy samples are indeed very different. As a result, an emulator (such as ours) trained on any single simulation (e.g., the SC SAM) would predictably perform poorer on an alternative simulation (e.g., TNG100-1) if the space of HODs covered by the test set lies outside of the space of HODs covered by the training set--such is the case that we have identified here. Improving the robustness of the emulator by training on a range of simulations, spanning a larger space of HODs, would likely improve our framework's generalizability.

\bibliography{references}
\bibliographystyle{aasjournal}

\end{document}